\documentclass[12pt]{iopart}

\usepackage{iopams}  
\usepackage{graphicx}
\usepackage{color}
\usepackage{pdfpages}
\def\beq{\begin{equation}}
\def\eeq{\end{equation}}

\begin{document}
\title[]{Quantum localization and electronic transport in covalently functionalized carbon nanotubes} 

\author{
Ghassen {Jema\"i}$^1$,
Jouda Jama {Khabthani}$^1$,
Guy {Trambly de Laissardi\`ere}$^2$,
Didier {Mayou}$^{3,4}$
}

\address{$^1$ 
Laboratoire de Physique de la Mati\`ere Condens\'ee, D\'epartement de Physique, Facult\'e des Sciences de Tunis, Universit\'e Tunis El Manar, Campus Universitaire, 1060 Tunis, Tunisia}
%\ead{ghassen.jemai@fst.utm.tn}
\address{$^2$ Laboratoire de Physique Th\'eorique et Mod\'elisation, CNRS and Universit\'e de Cergy-Pontoise, %2 av. A. Chauvin, 
95302 Cergy-Pontoise, France}
\address{$^3$ CNRS, Inst NEEL, F-38042 Grenoble, France.}
\address{$^4$ Universit\'e Grenoble Alpes, Inst NEEL, F-38042 Grenoble, France}

\ead{guy.trambly@u-cergy.fr}

\vspace{10pt}
\begin{indented}
\item[]October 2019
\end{indented}

\begin{abstract}
Carbon nanotubes are of central importance for applications in nano-electronics thanks to their exceptional transport properties.
They can be used as sensors, for example  in biological applications, provided that they are functionalized to detect specific molecules.
Due to their one-dimensional geometry the carbon nanotubes are very sensitive to the phenomenon of Anderson localization and it is therefore essential to know how the functionalization modifies their conduction properties and if they remain good conductors. Here we present a study of  the quantum localization induced by functionalization  in  metallic  single walled carbon nanotubes (SWCNT) with  circumferences up to 15\,nm. We consider  resonant and non-resonant adsorbates that represent two types of covalently functionalized groups with strong and moderate scattering properties. The present study provides a detailed analysis of the localization behaviour and shows that the localization length can decrease down to 20-50\,nm at concentrations of about 1 percent of adsorbates. On this basis we discuss the possible electronic transport mechanisms which can be either metallic like  or insulating like with variable range hopping.
\end{abstract}

\vspace{2pc}
\noindent{\it Keywords}: Carbon nanotube, functionalization, quantum transport

%Uncomment for Submitted to journal title message
\submitto{\JPCM}
%\maketitle

\section{Introduction}

Single walled carbon nanotubes (SWCNTs) are quasi-one-dimensional
cylindrical shape materials with $sp^{2}$ hybridized carbon atoms,
that exhibit unique physical and chemical properties. Since their
discovery in 1991, they have attracted much attention for their fundamental properties and their wide range of potential applications, such as for energy
storage \cite{yang_hydrogen_2019,muhulet_fundamentals_2018}, solar
cells \cite{fu_flexible-Solar-cells_2018, jeon_solar-cells_2018, oo_perovskite_2017}, 
nano-electronics and sensors devices \cite{yoo_Bacteria_2016, choi_sensor_Staphylococcus_2017}.

SWCNT get easily bundled forming agglomeration of nanotubes as a result of the strong Van der Walls energy interaction \cite{britz_noncovalent_2006}, and because of that it's always challenging to manipulate them efficiently for experiments. Yet surface functionalization allows to overcome this problem since ad-atoms and molecules covalently attached to it's surface can greatly reduce the effect of the strong Van der Walls interaction.
The functionalization also allows the compatibility toward other molecules to create new composite materials based on carbon nanotubes
\cite{khan_carbon_2016, kamran_functionalized_2019}. In particular, it has been reported
that SWCNT is ideal to create reliable and accurate electrochemical
biosensor, faster than conventional ones \cite{zhu_overview_Bio-sensor_2017,zaporotskova_review_sensor_2017,muhulet_fundamentals_2018,yoo_Bacteria_2016},
thanks to its low dimensionality, nanometric size, large surface area, high sensitivity and fast response. The two main components in such device, are the biological recognition element that detects a biological molecule and the transductor that collects a charge transferred from the biological molecule to convert it into an electrical signal unique to a specific detected element. 
Since a covalent functionalization can drastically alter the electronic properties of carbon nanotubes 
{%\textcolor{red}{
\cite{gomez-navarro_tuning_2005,Flores_2008}}, 
it can affect the detection process in a biosensor. 

{%\textcolor{red}{
Many theoretical analyses of electronic transport in SWCNT have been carried out using different numerical techniques, such as real space Kubo-Greenwood method for bulk systems \cite{Latil_2004,Latil2005,roche_inelastic_2010}, and Laudauer type formalisms for finite systems coupled to leads \cite{McEuen_1999,mceuen_electron_2004,Flores_2008,Li2009,Lopez-Bezanilla2009,KHOEINI_2009,ZIENERT_2013,Teichert_2014,TEICHERT_2017,Roche_2019-RevB}. Most of them discussed the importance of the Anderson localization due to charged defects or local defects such as vacancies, ad-atoms or ad-molecules. However, a global understanding of localization process based on numerical calculations that takes into account the effects of quantum interferences without approximations is still missing.
}
In this paper we investigate these effects for  SWCNT of diameter up to $4$-$5$\,nm   and large concentrations of functionalized sites up to 1 or a few percent. In particular we address the question of the coherent electronic transport in the nanotube and the occurence of localization effect which can diminish the conduction properties of the functionalized nanotube \cite{P-W_Anderson_1958, thouless_andersons_1970, thouless_localization-1D_1973}. Since strong localization effects are inherent to low dimensional conductors \cite{Roche_2019-RevB, nature_AL} it is important to have a clear description of its effect on these electronic devices.
{%\textcolor{red}{
Our approach allows to propose a detailed analysis of the conductivity and its correction due to quantum localization for a wide range of concentrations of resonant and non-resonant defects. 
It is confirmed that quantum localization is the key to understand the wide variety of transport regimes depending on the nanotube diameter, the adsorbates concentration and the type of functionalization. In particular we show that at low adsorbates concentration the system obeys the Thouless relation, typical of a one-dimensional system, 
which links the number of conducting channels, the elastic mean-free path and the localization length. The values of these fundamental lengths are estimated for resonant and non-resonant adsorbates with concentration ranging from 0.1\% to 4\%, which corresponds to different types of applications such as sensors. Yet we show also that these nanotubes can present a variety of other behaviors that emerge at sufficiently high concentration of adsorbates. For example we find that a transition form 1D localization behavior to 2D localization behavior can occur for resonant scatterers at a concentration of the order of 1\%. Also, depending on the adsorbate concentration and nanotube diameter we discuss the occurence of a regime of variable range hopping which could possibly be observed at room temperature. 
}

{%\textcolor{red}{
The paper is organized as follows. First section presents
the tight-binding model and the basic concepts used to analyze the quantum localization effect on transport. 
The numerical method to study SWCNT is presented Sec. \ref{sec_modeling} and in the supplementary material.
Then we analyze the localization in the cases of carbon nanotube functionalized with non-resonant (Sec. \ref{sec_transport_non-res}) or resonant  (Sec. \ref{sec_transport_res}) adsorbates with various concentrations in the range of 0.1\% to 4\%. 
A discussion is given Sec. \ref{sec_discussion} about the  transport mechanism metallic like or insulating like and in particular the possibility of variable range hopping.
}

\section{Model and Methodology}
\label{sec_modeling}

\subsection{Modeling resonant and non-resonants adsorbates in SWCNT}

A single walled carbon nanotube (SWCNT) is equivalent to a rolled
up stripe of a 2D graphene sheet.  The orientation of rolling
up graphene, is defined by the chiral vector $\vec{C}_{h}=n\vec{a}_{1}+m\vec{a}_{2}$,
where $\vec{a}_{1}$ and $\vec{a}_{2}$ are the unit cell vectors
of graphene, the $ n $ and $ m $ are coefficients used to define the chirality of the nanotube. The SWCNT diameter is $d=\frac{C_{h}}{\pi}$ where
$C_{h}=\left\Vert \vec{C}_{h}\right\Vert =a\sqrt{3(n^{2}+m^{2}+nm)}$
the circunference of the nanotube, with $a=0.142$\,nm the distance
between two neighboring carbon atoms. The ratio between the length and diameter $\frac{L}{d}\sim\left(\frac{length}{diameter}\right)$
can be as large as $10^{4}$ to $10^{5}$ \cite{saito_CNT_1998} and  carbon nanotube can be considered as a quasi-one-dimensional nano-structure.
In this  study the SWCNT has an infinite length which avoids edges effects. 

In this paper we consider the metallic zigzag single walled carbon
nanotube $\left(n,0\right)$ where the shape of the SWCNT cross section is zigzag and with ``$n$'' is a multiple integer of $3$. We took $n=30$, $n=60$ and  $n=90$. We show detailed results for $n=60$ in the article and compare them with the results for $n=30$ and  $n=90$ that are summarised  in the supplementary material.

We study the electronic properties of SWCNT using a first neighbour
tight-binding hamiltonian

\begin{equation}
\tilde{H}=-t\sum_{i,j}\left(c_{i}^{\dagger}c_{j}+c_{i}c_{j}^{\dagger}\right),
\label{H}
\end{equation}
where $c_{i}^{\dagger}\left(c_{i}\right)$ is the creation (annihilation) operator. The hopping integral $ t=2.7$\,eV allows the electron to hop from an atom site to one of its three first neighbours. We consider that all the onsite energies are the same and they are all set to
zero $(\varepsilon_{0}=0$\,eV). 

When an ad-atom or molecule is attached to the nanotube surface it creates an adsorbate and a covalent bond. This covalent bond is equivalent to taking out a $p_{z}$ orbital. It has been established that adsorption on top of a carbon atom  allows to keep the $sp^{2}$ hybridization \cite{pereira_modeling-disorder_2008,setaro_preserving_pi_2017}.
Therefore we model adsorption on top of a carbon site by simply removing the corresponding $p_{z}$ orbital in equation (\ref{H}).
We therefore assume that an adsorbate can be seen as a carbon vacancy present in the SWCNT without disturbing the other carbon atoms positions. Note that an adsorbate is a local defect and produces inter-valley scattering which leads to
Anderson localization \cite{Harju_2D-AL_2014, Harju_AL_crossover_2013, Harju_linear-scaling_2014}. 

We consider two types of local defects obtained by removing one isolated $p_{z}$ orbital or two neighboring $p_{z}$ orbitals. These two types of defects correspond respectively to resonant and non-resonant adsorbates. Indeed the effect of an isolated local defect is entirely  described by an energy dependent T-matrix  $ T(E) $. Resonant adsorbates are modeled by removing one isolated $p_{z}$ orbitals and the corresponding T-matrix presents a strong resonance  at the charge neutrality point (CNP) creating a  peak in the density of states around the CNP. A similar peaks exist in graphene for which $ T(E) $ even presents a divergence at the CNP.
Non-resonant adsorbates, are modelled by removing two neighbouring
orbitals at the same time. In that case the T-matrix   varies smoothly with energy and does not show resonance close to the CNP. Therefore no peak appears  in the DOS around the CNP. Resonant and non-resonant adsorbates affect differently the electronic
transport, and in particular the resonant adsorbates produce a stronger scattering than the non-resonants adsorbates close to the CNP. 

\subsection{Analysis of Quantum Localization }

%The electrical conductivity is obtained using the Einstein
%relation: 

%\begin{equation}
%\sigma\left(E\right)=e^{2}n\left(E\right)D\left(E\right)
%\end{equation}

%In the diffusive regime and without quantum interference when $l_{e}$ and $L_{i}$ are comparable the electrical conductivity is maximum this maximum
%value $\sigma_{M}=e^{2}n\left(E\right)D_{M}$. 

{%\textcolor{red}{
Following the real space Kubo-Greenwood method \cite{mayou_calculation_1988,mayou_real-space_1995,roche_Quasiperiodic_1997,Roche99,Triozon02, mayou_quasi-cristals_2007} 
for
transport properties of electron in the nanotube, we compute  $X^2(E,t)$ which is the average square of the quantum spreading along the direction of the nanotube,  after a time $t$ and for states at energy $E$. 
That method has been also used to study
graphene monolayer \cite{trambly_graphene_2013}, bilayer \cite{missaoui_mobility_2018,missaoui_numerical_2017}.
}
For an electron in a 1D system propagating in a static disordered potential  we know that all states are localized and quantum diffusion is limited by the localization length $\xi (E)$. Yet the time evolution of $X^2(E,t)$ contains much information about the localization phenomena. In order to do this analysis we simulate how  inelastic scattering affects diffusion and conductivity.
 At the simplest level the effect of inelastic scattering is described by introducing an inelastic scattering time $\tau_{ie}$ and a time of phase coherence of the wave-function  $\tau_{\Phi}$. These two times can be different and their relative value can depend on the studied system. In carbon nanotubes it has been argued that $\tau_{\Phi}$ can be smaller than $\tau_{ie}$ because loss of phase coherence can be due predominantly to long wavelength acoustic modes which have a small effect on $\tau_{ie}$   
\cite{roche_inelastic_2010}. In the present study we use the relaxation time approximation (RTA) which consist in assuming that the velocity correlation function of the system is destroyed by inelastic scattering and the associated loss of wave-function coherence. In this picture their is no distinction between $\tau_{\Phi}$ and  $\tau_{ie}$. We therefore introduce a parameter $\tau_{i}$ such that $\tau_{i}\simeq \tau_{ie}\simeq\tau_{\Phi}$. 
Associated to  $\tau_{i}$ there is a length scale called the inelastic mean-free path $L_{i}(E,\tau_{i})$ which is the typical propagation length at energy $E$ after a time $\tau_{i}$ (see supplementary material for precise definition). Beyond $\tau_{i}$ the transport regime is diffusive because of the loss of coherence, and  the diffusivity is given by $D(E,\tau_{i})=\frac{L_{i}^{2}\left(E,\tau_{i}\right)}{\tau_{i}}$. The expression of electrical conductivity becomes
\begin{equation}
\sigma\left(E_{F},\tau_{i}\right)=2e^{2}n\left(E_{F}\right)D\left(E_{F},\tau_{i}\right),
\label{sigma1}
\end{equation}
where $n(E_{F})$ is the density of states per unit length (per spin) at the Fermi energy $E_{F}$.
As shown below the variation of the conductivity $\sigma\left(E_{F},\tau_{i}\right)$ with $L_{i}(E,\tau_{i})$  or with $\tau_{i}$  contains essential information about the quantum transport. It allows to extract also the elastic mean-free path $l_e(E)$ and the localization length $\xi(E)$, and will be discussed in detail.  For example an important quantity is $\sigma_{M}(E_{F})$ the maximum of the conductivity as a function of $L_{i}(E,\tau_{i})$. In the following $\sigma_{M}(E_{F})$ is named the microscopic conductivity. It corresponds to a situation where the inelastic mean-free path is of the order of the elastic mean-free path so that the effect of quantum localization cannot develop and decrease the conductivity. $\sigma_{M}(E_{F})$  is given in term of the mean-free path $l_e(E)$ by
\begin{equation}
\sigma_{M}(E_{F})=G_{0} N_{ch}(E_{F}) l_e(E_{F}),
\label{sigma2}
\end{equation}
where $G_{0}=2e^2/h$ is the quantum of conductance and $N_{ch}(E)=n(E)/n_{0}$ is  the ratio between the density of states per unit length and per spin $n(E)$ at the energy $E$ and the density of states for one channel $n_{0}=2/hV_{G}$ where $V_{G}$ is the velocity in graphene at the CNP. $n_{0}$ is  half the density of states at the zero energy in the nanotube (there are two conduction channels near $E=0$ one close to each Dirac point). Therefore  $N_{ch}(E)$ is close to the number of channel at the energy $E$ but tends to be higher because the contribution to the density of states of a channel close to its minimum energy  (Van Hove singularity) is higher than $n_{0}=2/hV_{G}$.

\section{Electronic transport for non-resonant adsorbates}
\label{sec_transport_non-res}

\begin{figure}
    \centering
    \includegraphics[height=5cm,width=5cm]{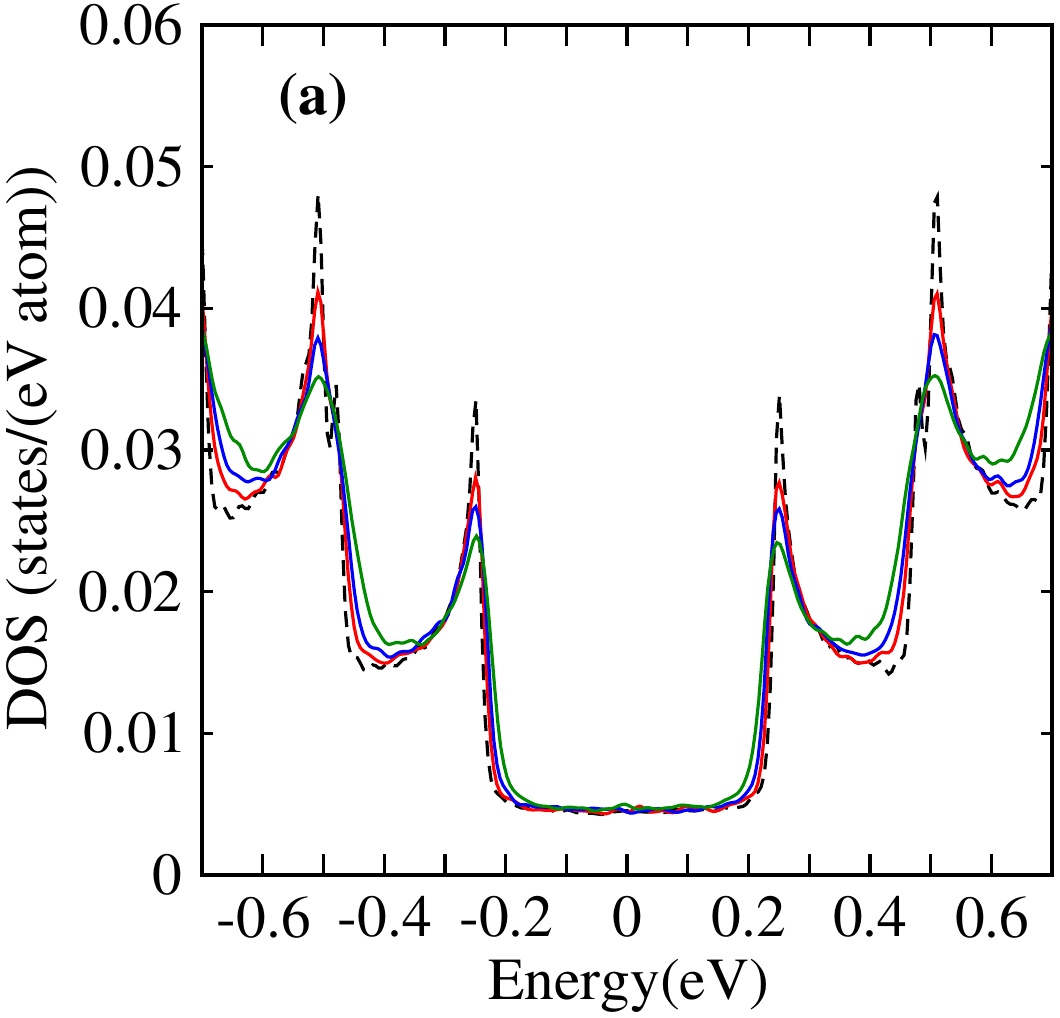} 
    \includegraphics[height=5cm,width=5cm]{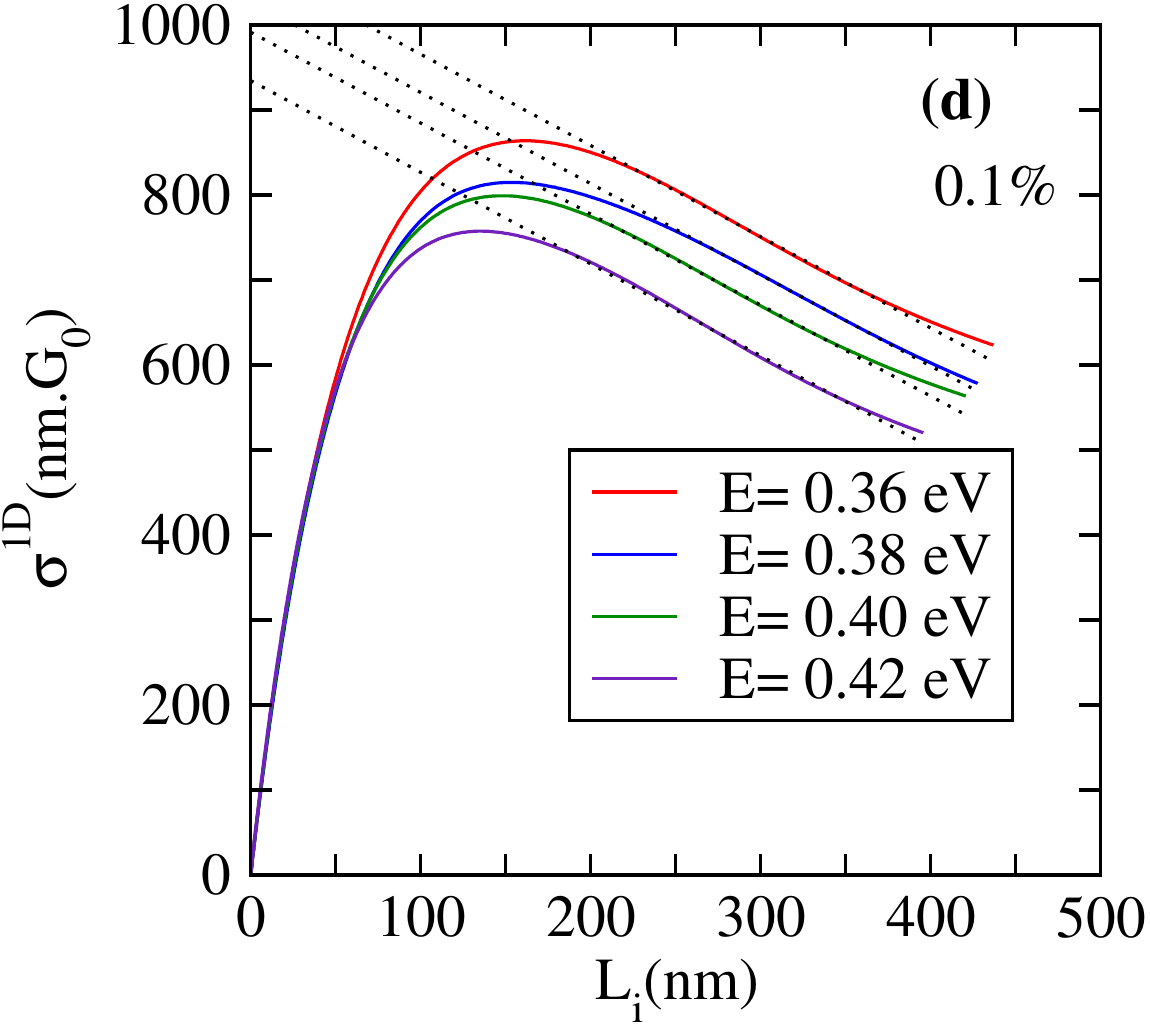}\\
	\includegraphics[height=5cm,width=5cm]{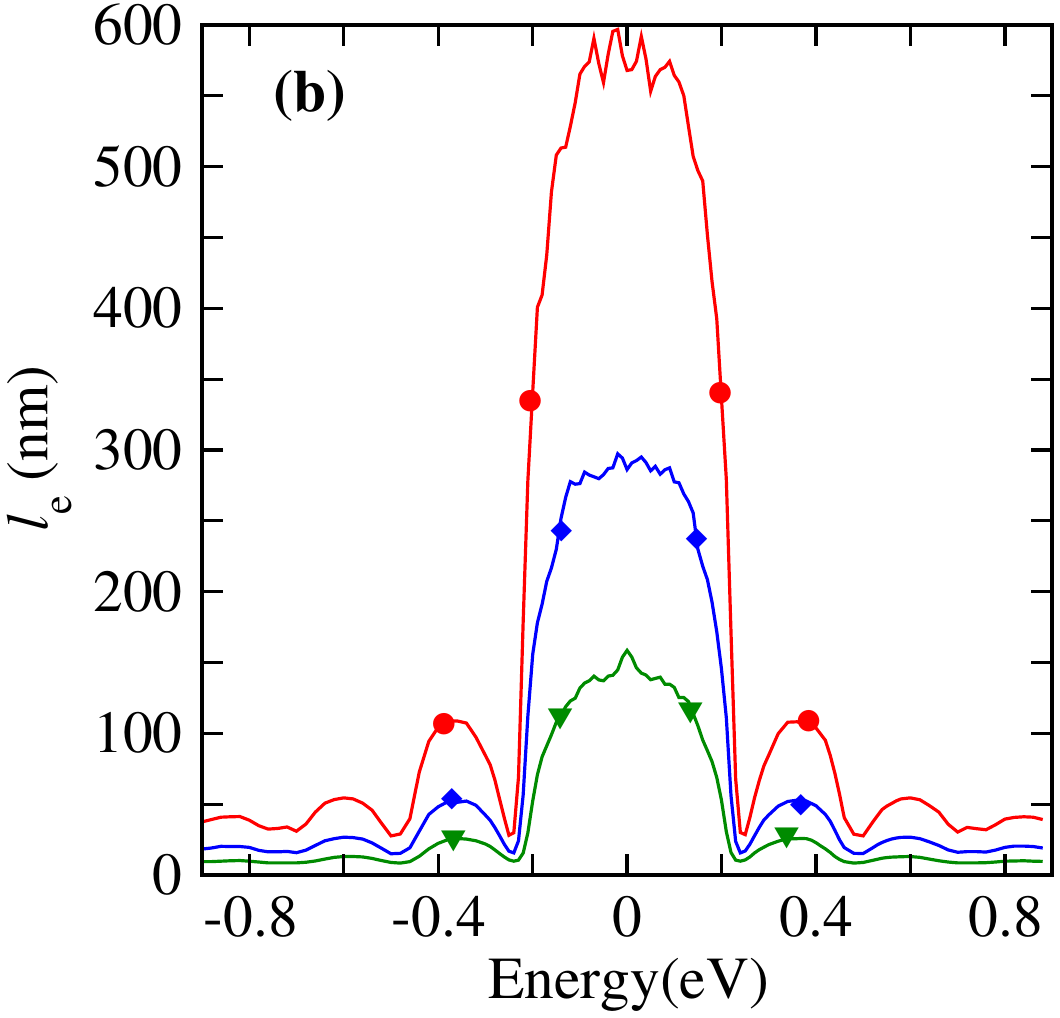}
	\includegraphics[height=5cm,width=5cm]{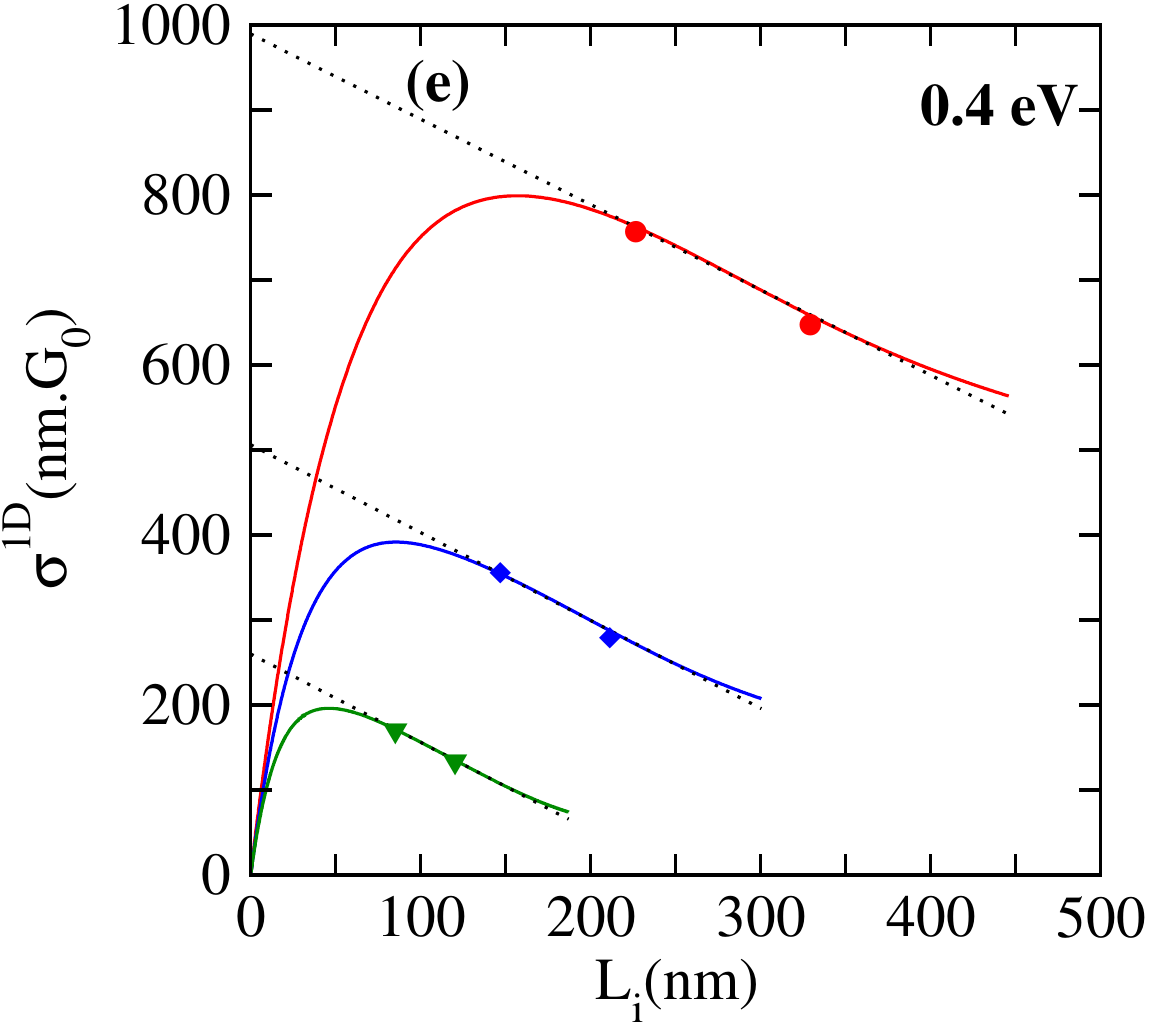}\\
    \includegraphics[height=5cm,width=5cm]{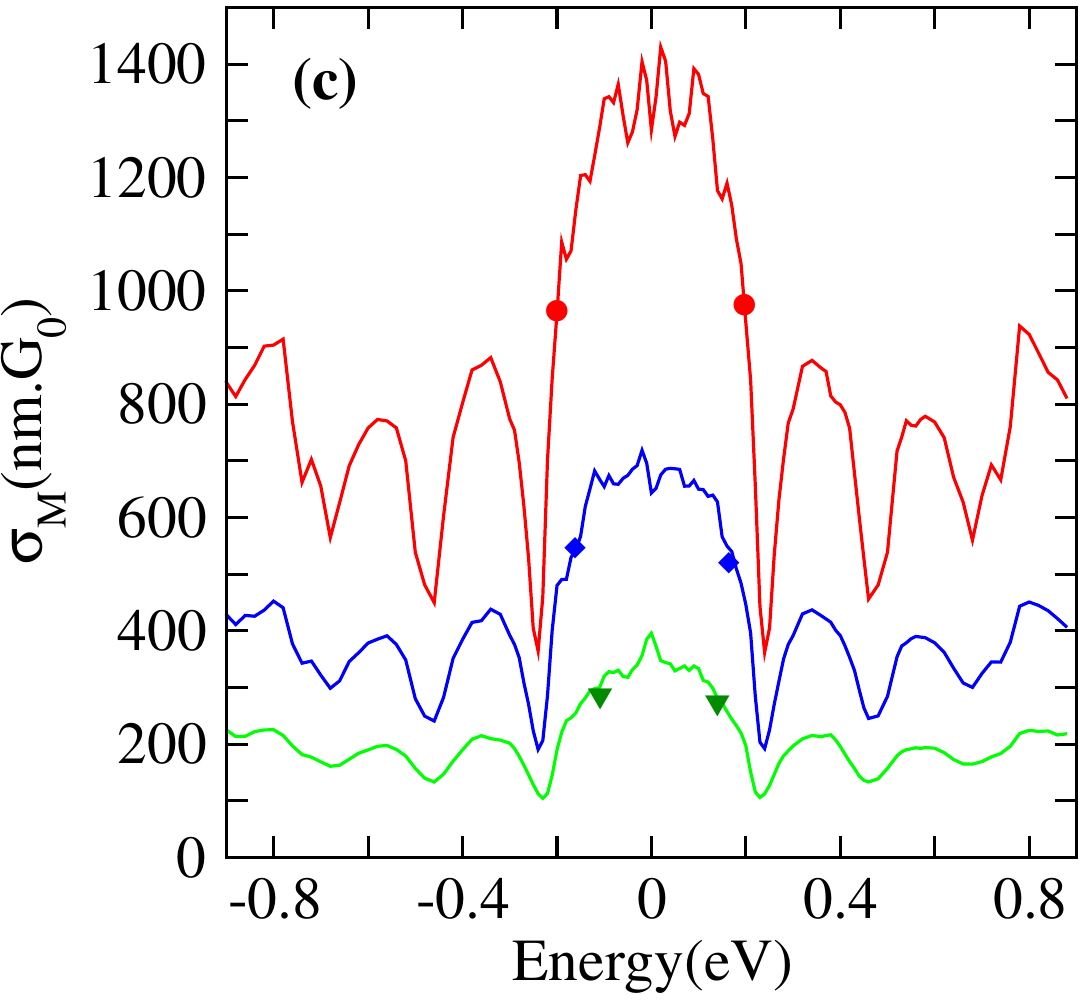}
     \includegraphics[height=5cm,width=5cm]{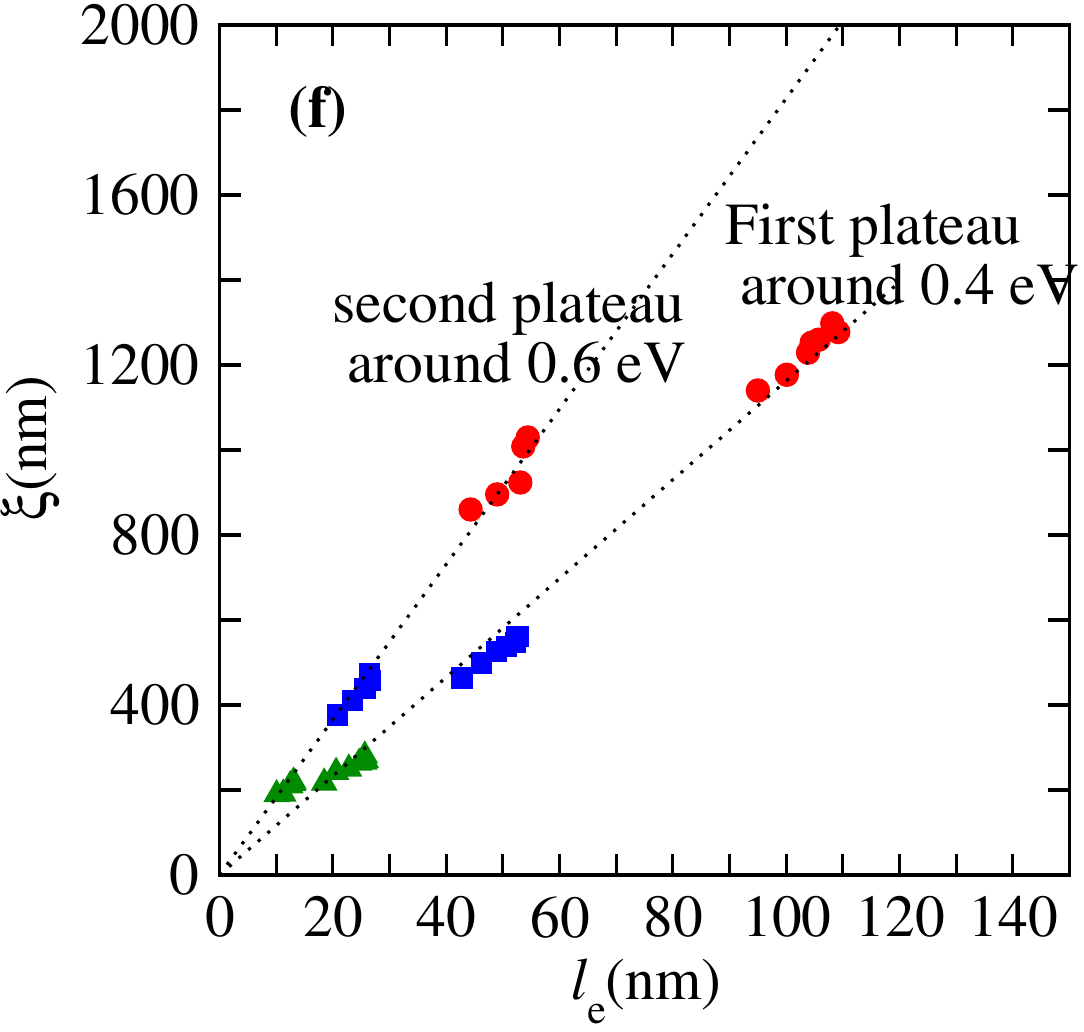}
\caption{Low concentration of non-resonant adsorbates: $0.1$\% (red/filled circle), $0.2$\% (blue/filled square) and $0.4$\% (green/filled triangle). (a) Density of states versus energy, (b) Elastic mean-free path $ l_{e} $ versus energy, (c) Microscopic conductivity $ \sigma_{M} $ versus energy, (d) Electronic conductivity $ \sigma $ versus inelastic mean-free path $ L_{i} $ for 0.1\% of non-resonant adsorbates, (e) Electronic conductivity $ \sigma $ versus inelastic mean-free path $ L_{i} $ at $E=0.4$\,eV, and (f) Localization length $ \xi $ versus elastic mean-free path $ l_{e} $ (Thouless relation).}
\label{fig_low_non-resonant}     
\end{figure}

\begin{figure}

    \centering
    \includegraphics[height=5cm,width=5cm]{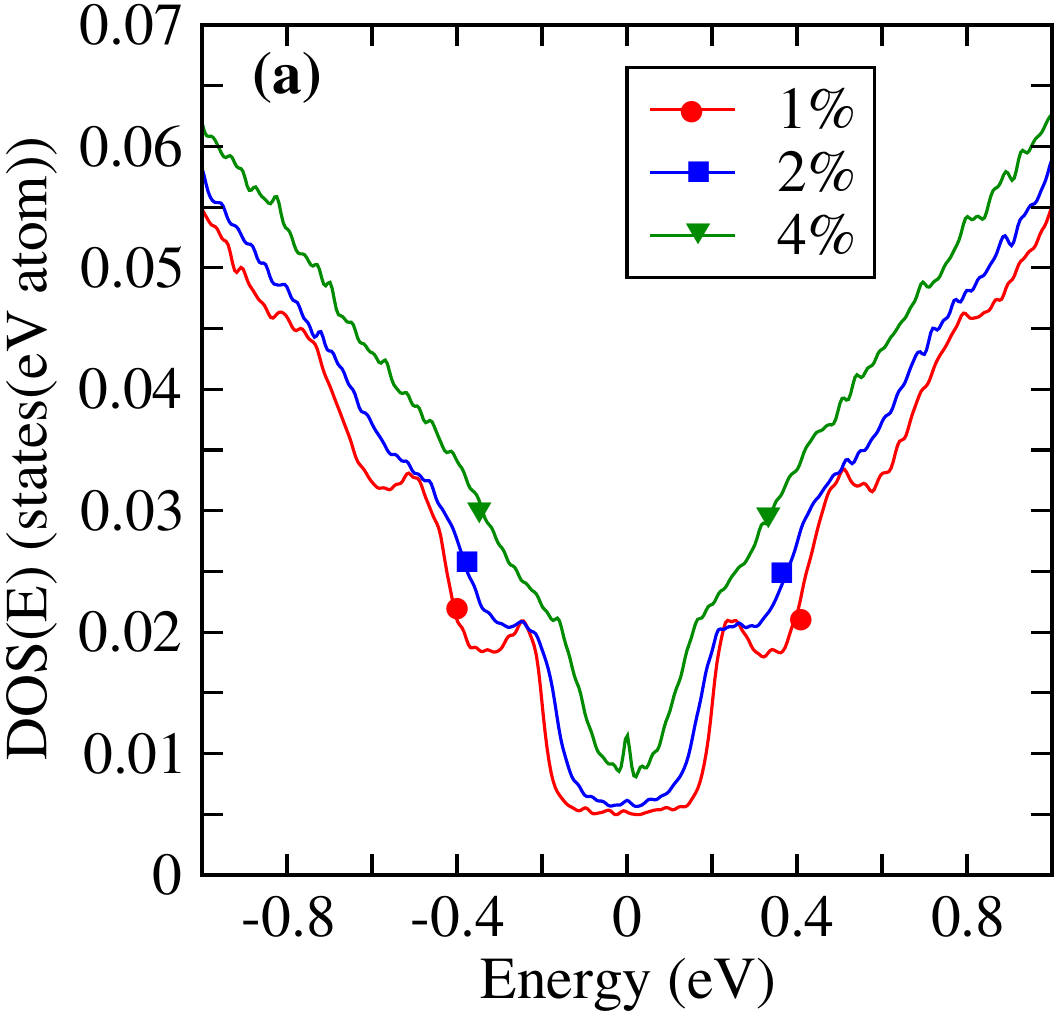} 
    \includegraphics[height=5cm,width=5cm]{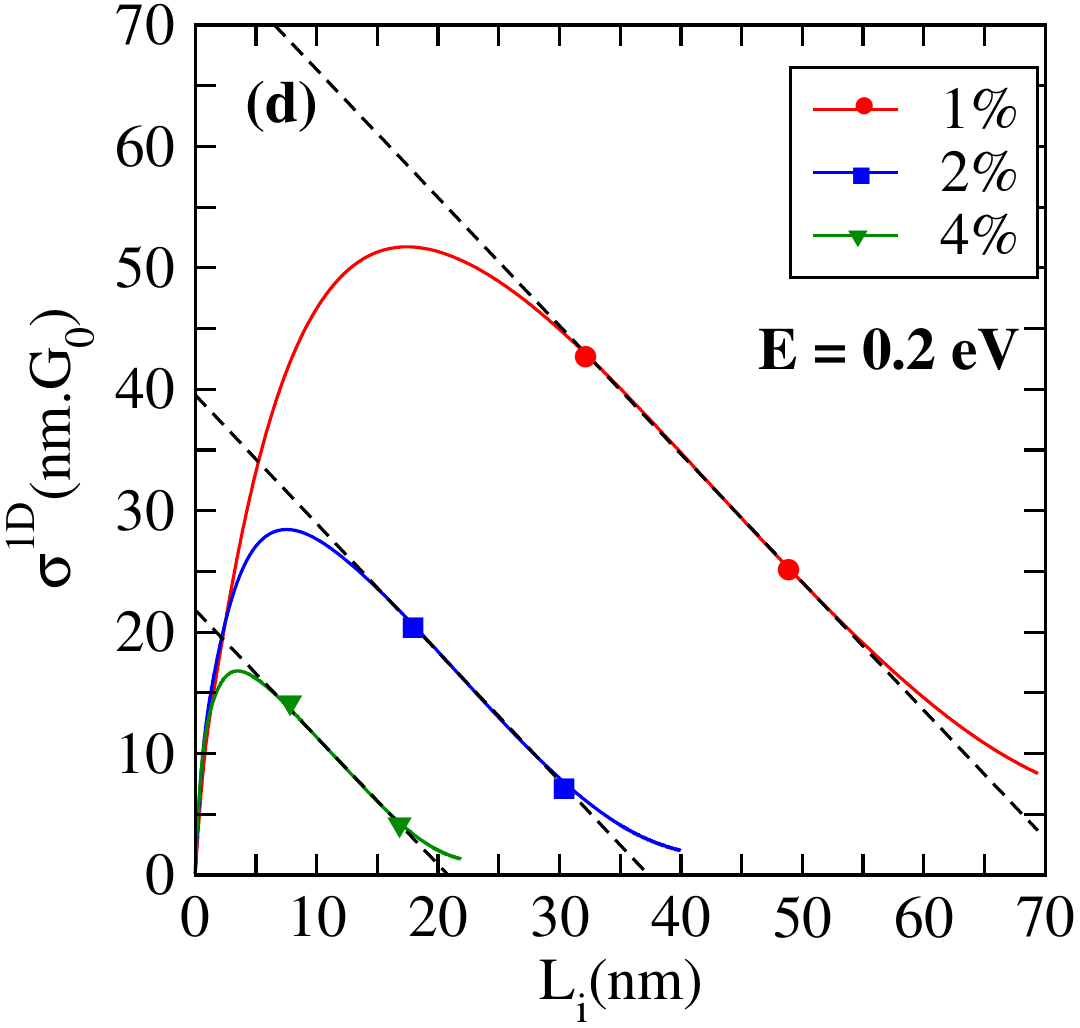}\\
    \includegraphics[height=5cm,width=5cm]{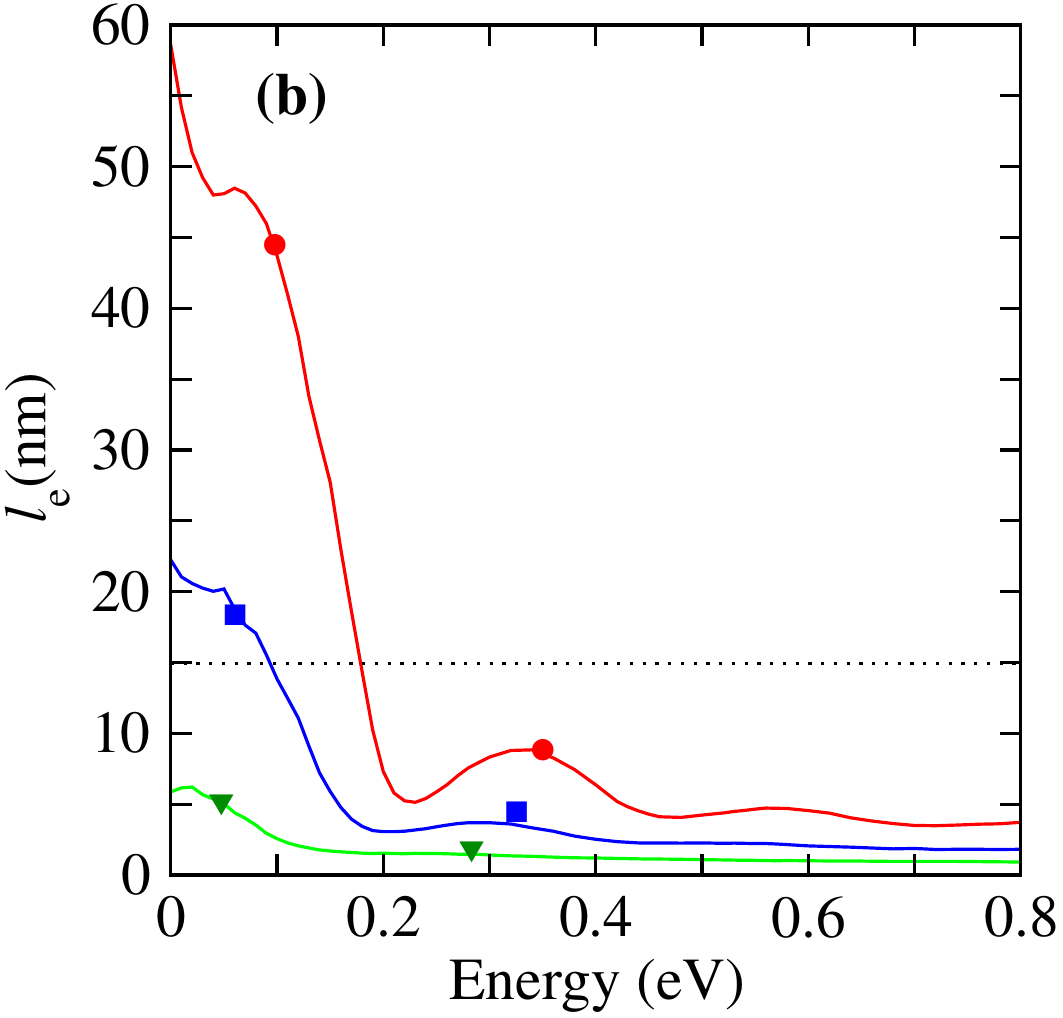}
    \includegraphics[height=5cm,width=5cm]{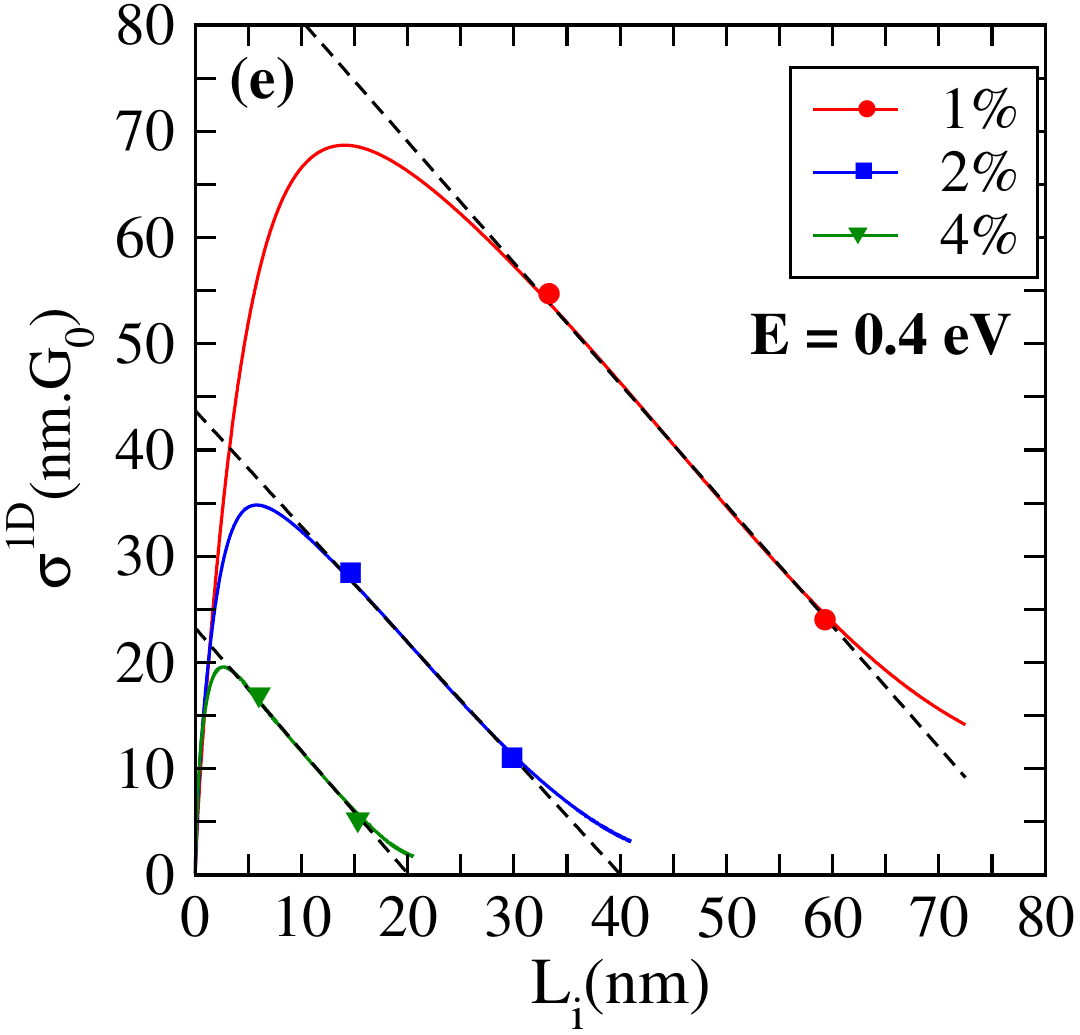}\\
     \includegraphics[height=5cm,width=5cm]{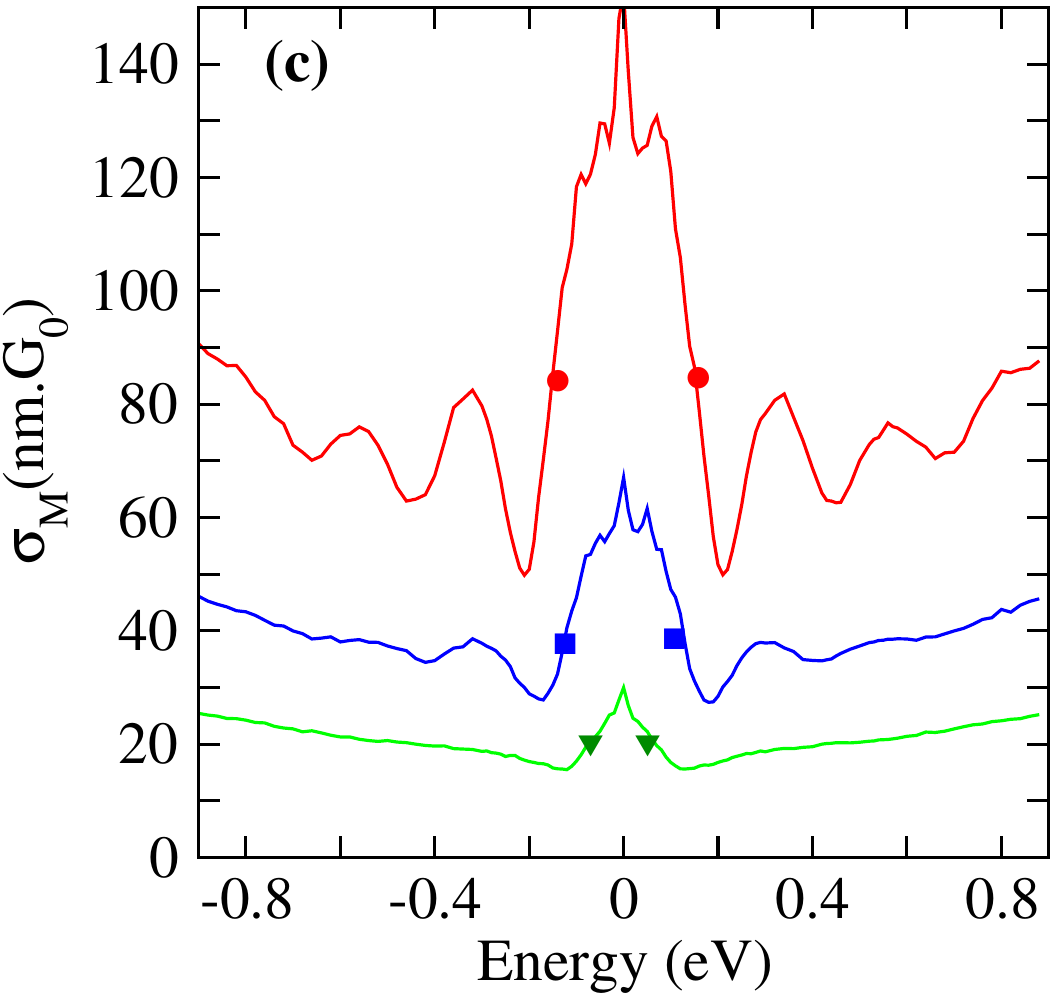}
     \includegraphics[height=5cm,width=5cm]{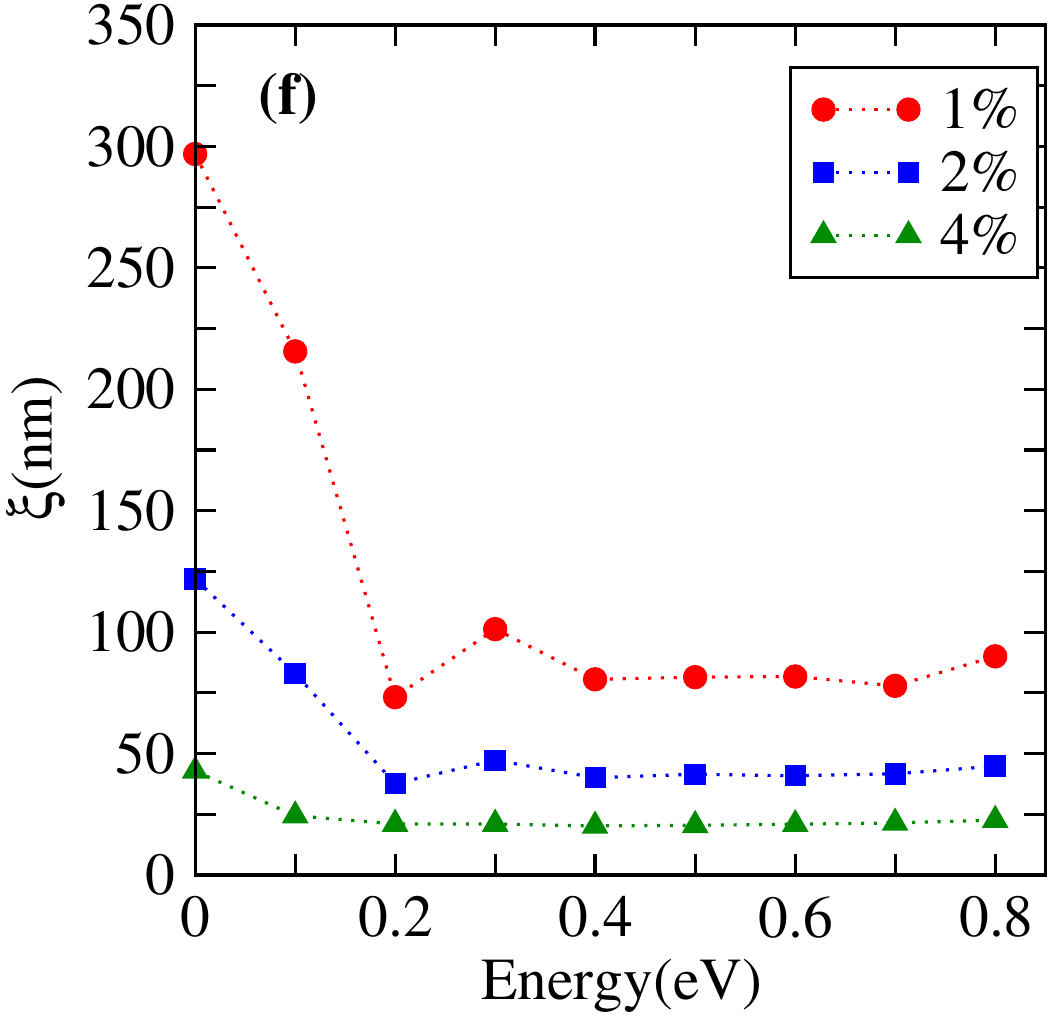}
\caption{High concentration of non-resonant adsorbates: 1\% (red/filled circle), 2\% (blue/filled square) and 4\% (green/filled triangle). (a) Density of states versus energy, (b) Elastic mean-free path $ l_{e} $ versus energy, (c) Microscopic conductivity per unit surface $ \sigma_{M}/C_{h} $ versus energy, (d) Electronic conductivity $ \sigma $ versus inelastic mean-free path $ L_{i} $ at $E=0.2$\,eV, (e) Electronic conductivity $ \sigma $ versus inelastic mean-free path $ L_{i} $ at $E=0.4$\,eV, and (f) Localization length $ \xi $ versus energy.}
\label{fig_high_non-resonant}     
\end{figure}

We investigate the electronic behavior of metallic SWCNT $\left(60,0\right)$
with a large diameter $d=4.75$\,nm and a circumference $C_{h}=14.94$\,nm,
in presence of non-resonant adsorbates. We consider first the low concentration cases 
(0.1\%, 0.2\% and 0.4\%) (Fig. \ref{fig_low_non-resonant}) and then higher concentrations cases (1\%, 2\% and 4\%) (Fig. \ref{fig_high_non-resonant}).

Figure \ref{fig_low_non-resonant}-a shows the total density of
states versus energy. For low energy near the CNP, 
the DOS is weakly affected by disorder. At larger energy  and in particular close to Van Hove singularity (VHS), the modification of the  DOS is more important.

Figure \ref{fig_low_non-resonant}-b, shows the elastic mean-free
path $l_{e}$ as a function of energy $E$.
$l_{e}$ is maximum around the CNP energy and decreases by crossing a VHS. It reaches a new maximum in between two successive VHS.  Indeed  the probability of
scattering increases as a function of energy according to the DOS and therefore the mean scattering time and the mean-free path decrease. In addition the new channel that appears at a VHS has a low band velocity. %at the VHS.
The probability of scattering also increases by increasing the concentration of adsorbates which results in a shorter elastic mean-free path. The results show that $l_{e}$ is inversely proportional to the concentration of adsorbates, which is in accordance with the Fermi Golden rule.

The microscopic conductivity $\sigma_{M}$ versus energy is shown in Fig.
\ref{fig_low_non-resonant}-c. This quantity is given by equation (\ref{sigma2}) and is proportional to  the product of results in Figs. \ref{fig_low_non-resonant}-a  and \ref{fig_low_non-resonant}-b. The evolution of $\sigma_{M}$ as a function of energy,
follows globally the  behaviour of $l_{e}$ as a function of energy. 
As for the elastic mean-free path the increase of concentration of adsorbates decreases the conductivity
$\sigma_{M}$ which is inversely proportional to the concentration of adsorbates. $\sigma_{M}$ is
a semi-classical quantity corresponding to the diffusive regime when $L_{i}$ and
$l_{e}$ are comparable. Thus, it does not depend on the quantum
localization effects. 

Figures \ref{fig_low_non-resonant}-d  and \ref{fig_low_non-resonant}-e present the electronic conductivity as a function of the inelastic mean-free path. 
At low $L_{i}$, %that
  $\sigma(E_{F},\tau_{i})$  increases in the ballistic regime. Then it reaches its maximum and for large $L_{i}$ decreases because of Anderson localization. In the localization regime, we find that the conductivity behaves according to the 1D scale
dependent conductivity \cite{Ramakrishnan_1985} given by
\begin{equation}
\sigma (E_{F},L_{i})=\sigma_{M}(E_{F}) - G_{0}\left(L_{i}-L_{e}\right).
\label{eq3}
\end{equation}
Here $L_{e}$ is defined as the value of $L_{i}$ for which the extension of the linear regime crosses the maximum value of $\sigma_{M}(E_{F})$. $L_{e}$  is expected to be of the order of the elastic mean-free path $l_{e}$  and indeed we find numerically that $L_{e}\simeq 2 l_{e}$. The behavior given by equation (\ref{eq3}) is consistent with the standard theory of localization in one dimensional conductors \cite{Ramakrishnan_1985}.
%%%%%%%%%%%%%%%%

%%%%%%%%%%%%%%%%
%%%%%%%%%%%%%%%%
%%%%%%%%%%%%%%%%
%%%%%%%%%%%%%%%%%

%In Figure \ref{fig_1}-d and Figure \ref{fig_2}-d, we show the electronic conductivity as a function of inelastic mean
%free path $\sigma^{1D}\left(L_{i}\right)$ with 0.1\% of non-resonant
%and resonant adsorbates. At first for a very short $L_{i}$, $\sigma^{1D}$evolves
%ballistically until reaching a maximum value in the diffusive regime.
%For a large $L_{i}$ however, $\sigma^{1D}$ decreases towards the
%localisation regime. It is also shown that with non-resonant adsorbates,
%$\sigma^{1D}\left(L_{i}\right)$ reaches the localisation regime at
%the energy E=0.42 eV faster than for E =0.36 eV, however with resonant
%adsorbates $\sigma^{1D}\left(L_{i}\right)$ reaches the localisation
%regime for lower energies (E=0.36 eV).faster than for higher energies
%(E=0.42 eV). So the localisation length $\xi$ decreases as a function
%of energy for non-resonant adsorbates and increases as a function
%of energy for resonant adsorbates. Furthermore, Figure \ref{fig_1}-e
%and Figure \ref{fig_2}-e, shows that an increase
%of adsorbates concentration lead to a shorter localisation length

The localization length $\xi(E)$ is defined as the value of $L_{i}$ for which the extrapolation of equation (\ref{eq3}) cancels. Therefore using equation (\ref{sigma2}) and since $L_{e}\simeq 2 l_{e}$ one get
\begin{equation}
\xi(E)\simeq(N_{ch}(E)+2)l_{e}(E),
\label{Th}
\end{equation}
where, in this Thouless relation \cite{beenakker_random-matrix_thouless_1997},
$N_{ch}(E)=n(E)/n_{0}$. As explained previously  $N_{ch}(E)$ is close to the number of channels at the energy $E$ but tends to be higher. As shown in the Fig. \ref{fig_low_non-resonant}-f, the localization length $\xi$
evolves as a linear function of elastic mean-free path $l_{e}$ as
predicted by the Thouless relation with a slope that increases with the number
of conduction channels $N_{ch}$. For the first
plateau around 0.4 eV, $N_{ch}\approx 6$  and $N_{ch}\approx 10$ for the second plateau
around 0.6 eV. These results confirm the accuracy of our numerical calculations.

Figure \ref{fig_high_non-resonant}  presents a similar study as Fig. \ref{fig_low_non-resonant} but for higher concentrations (1\%, 2\% and 4\%), the modification of the DOS by disorder becomes more important and the VHS are smeared out. In addition for the concentration (4\%) the density of states is strongly modified even close to the CNP. Despite these differences the variations of $l_{e}(E)$, $\xi(E)$ and $\sigma_{M}(E)$ remain similar and are still  inversely proportional to the concentration. The linear variation of conductivity with $L_i$ is also given  by equation (\ref{eq3}). Note that even at high concentration of non-resonant adsorbates the localization length is allways larger than the circunference of the nanotube $ \xi>C_{h} $ as shown in Fig. \ref{fig_high_non-resonant}-f. 

We conclude therefore that for the non-resonant adsorbates the nanotube behaves as a 1D conductor with a relatively small amount of disorder, for which the microscopic conductivity $\sigma_{M}(E)$ and the elastic mean-free path  $l_{e}(E)$  are obtained from the Fermi golden rule. In addition our results confirm the Thouless relation between the localization length $\xi(E)$,  the elastic mean-free path  and the number of conduction channels.

\section{Electronic transport for resonant adsorbates }
\label{sec_transport_res}

%%%%%%%%%%%%%%%%
\begin{figure}
    \centering
    \includegraphics[height=5cm,width=5cm]{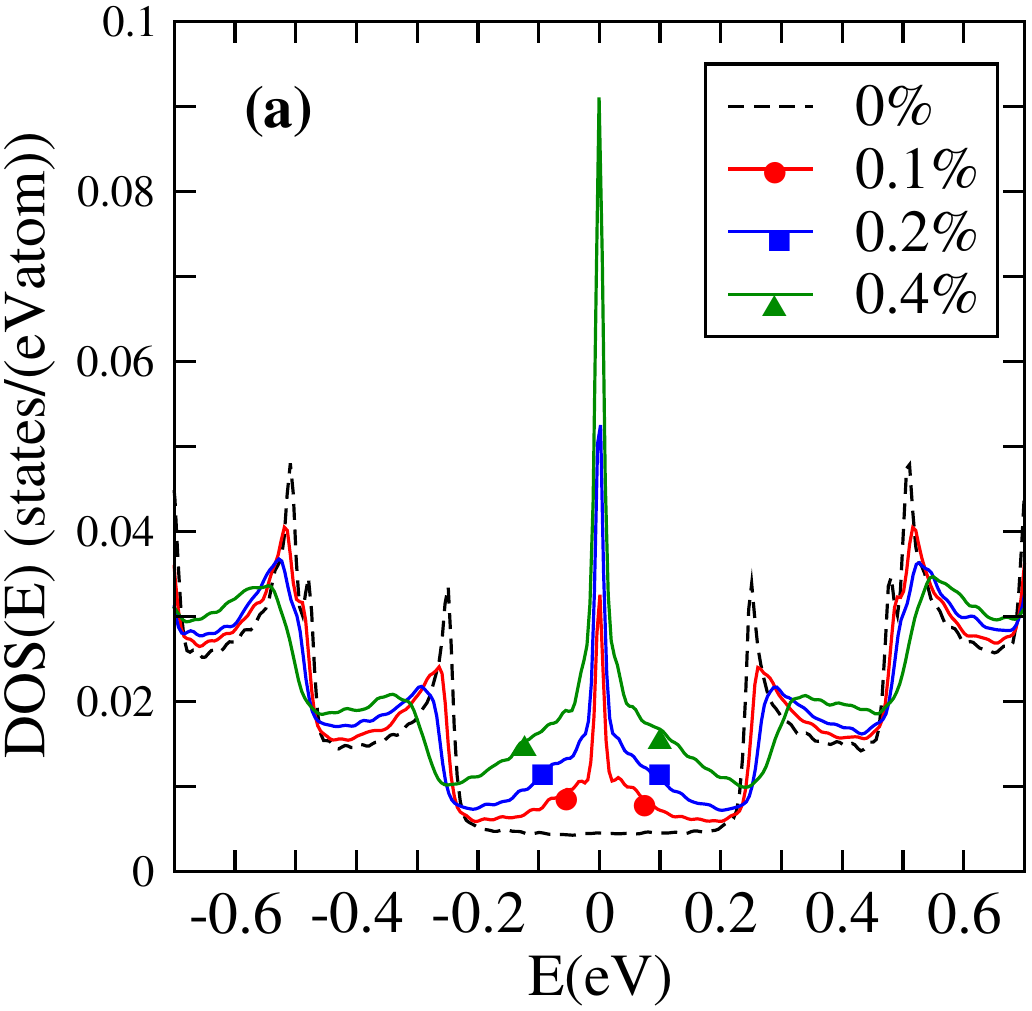} 
    \includegraphics[height=5cm,width=5cm]{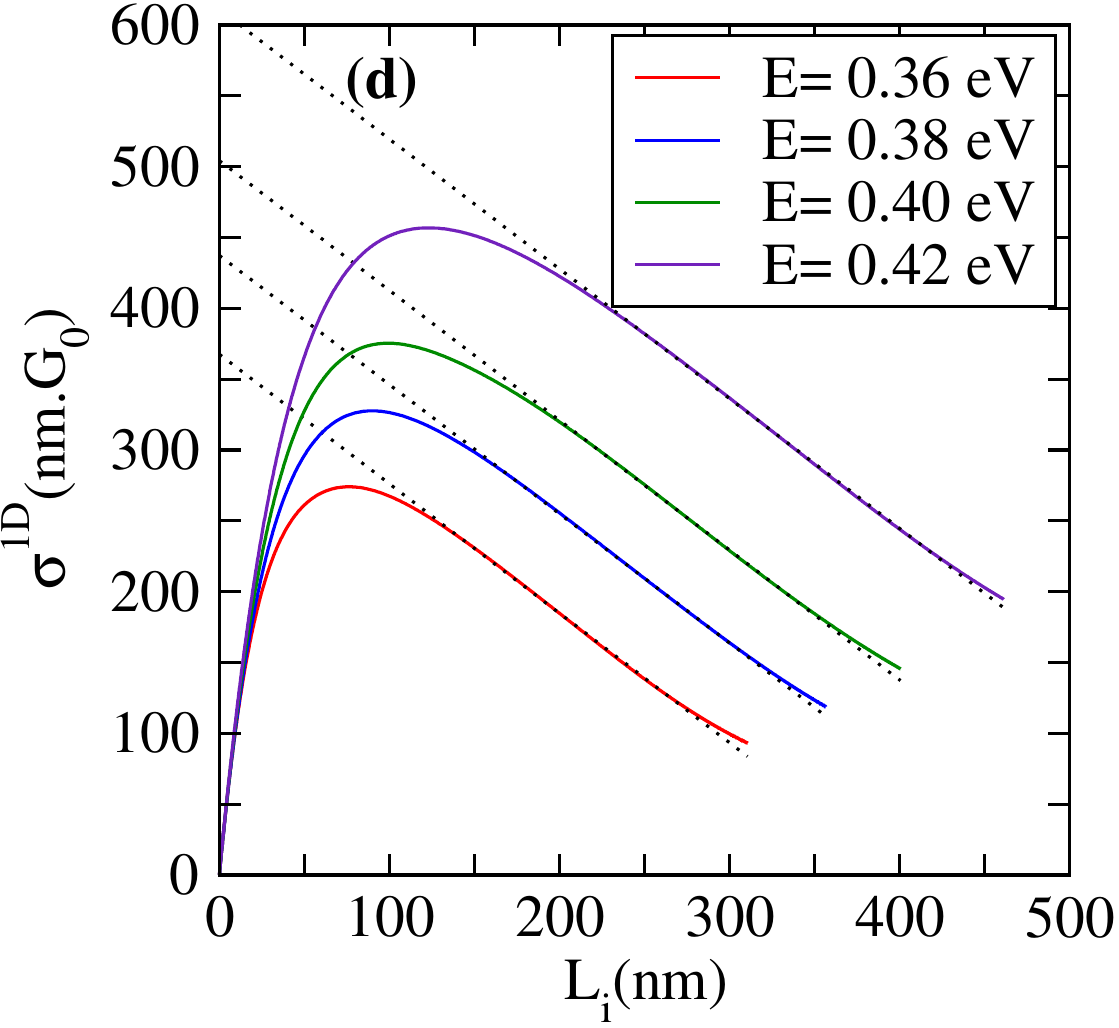}\\
    \includegraphics[height=5cm,width=5cm]{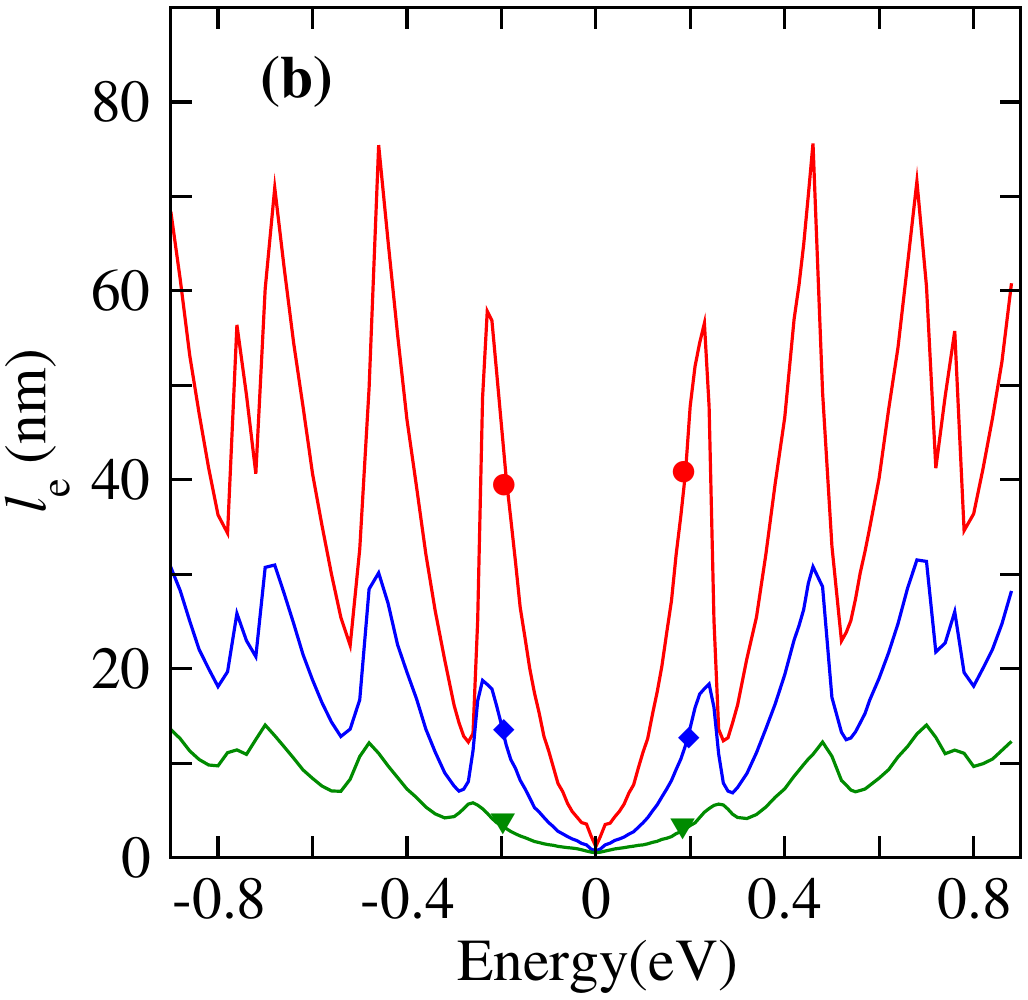}
    \includegraphics[height=5cm,width=5cm]{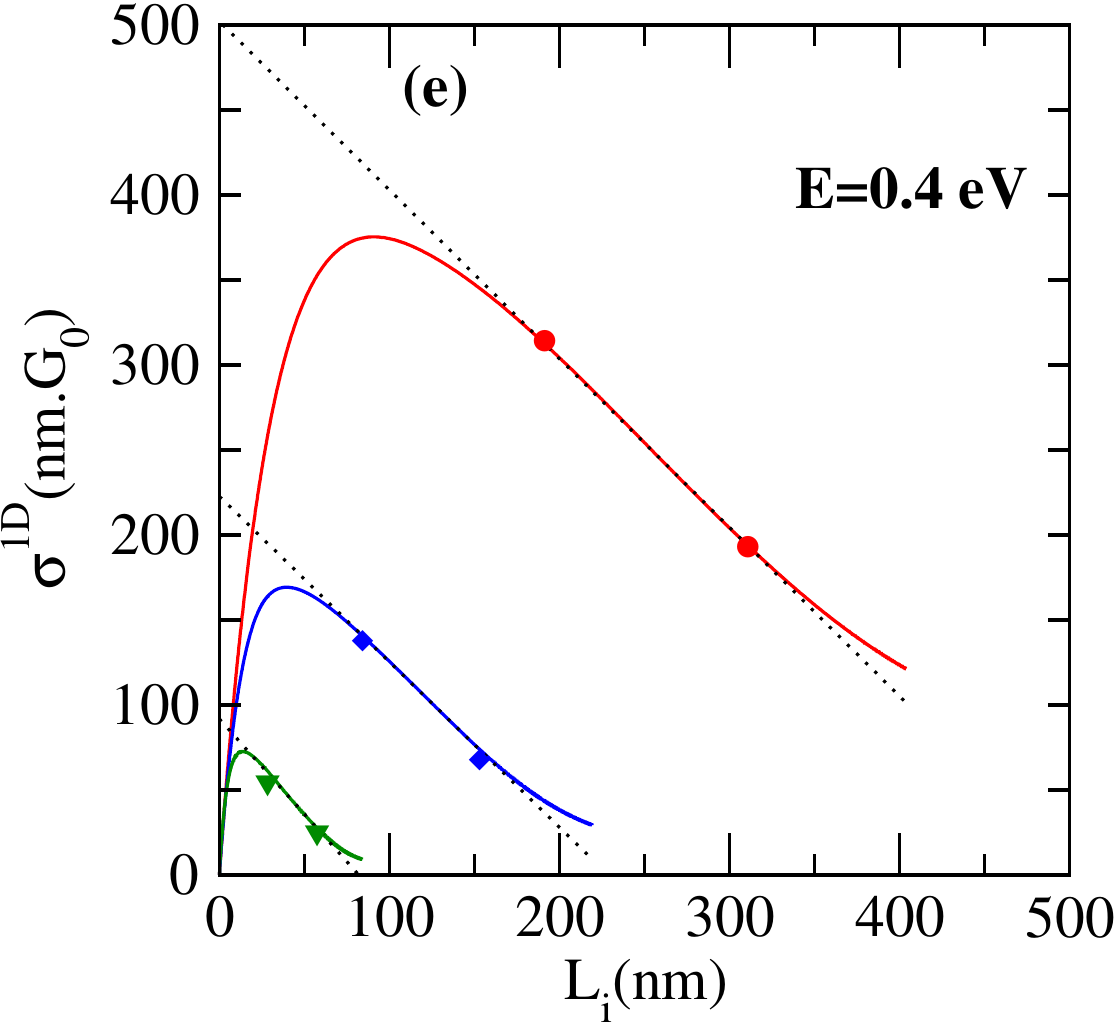}\\
     \includegraphics[height=5cm,width=5cm]{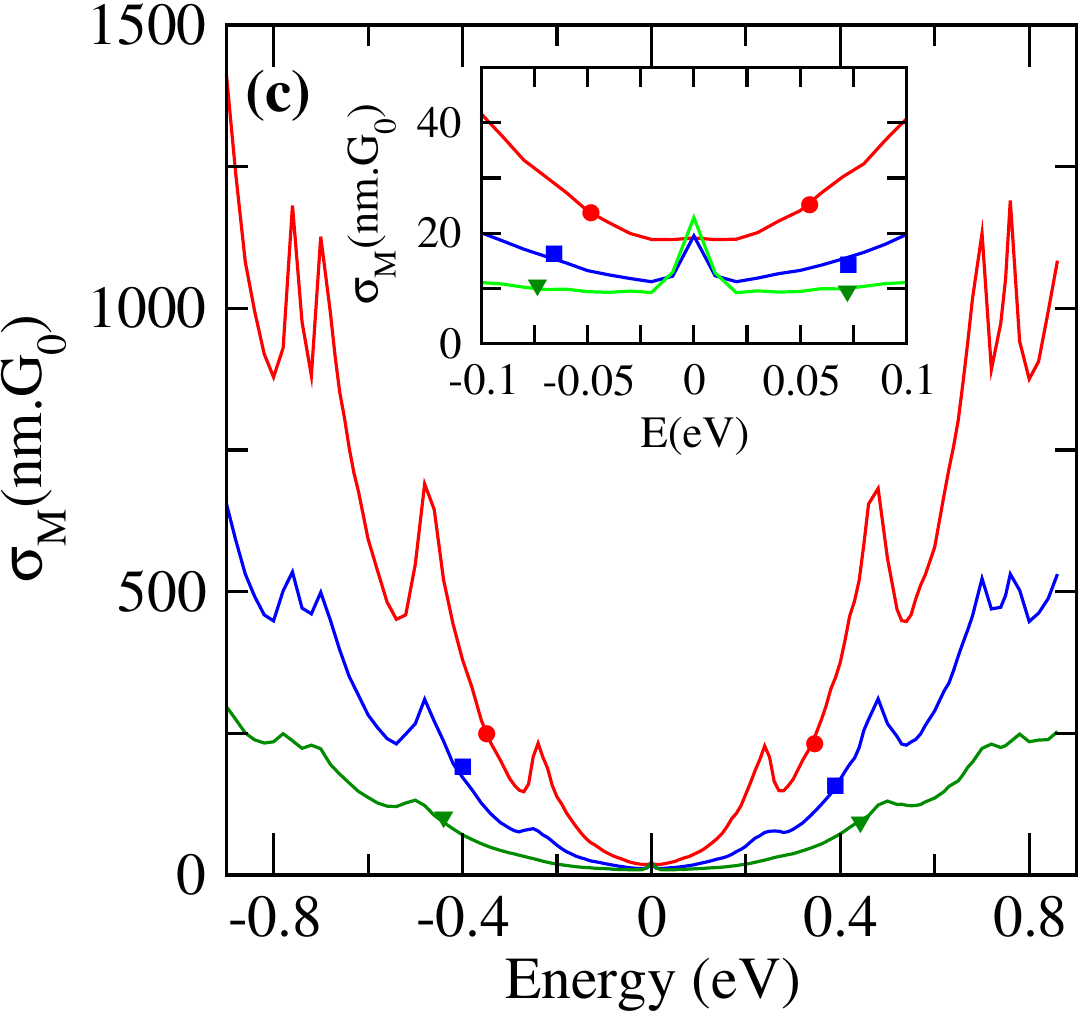}
     \includegraphics[height=5cm,width=5cm]{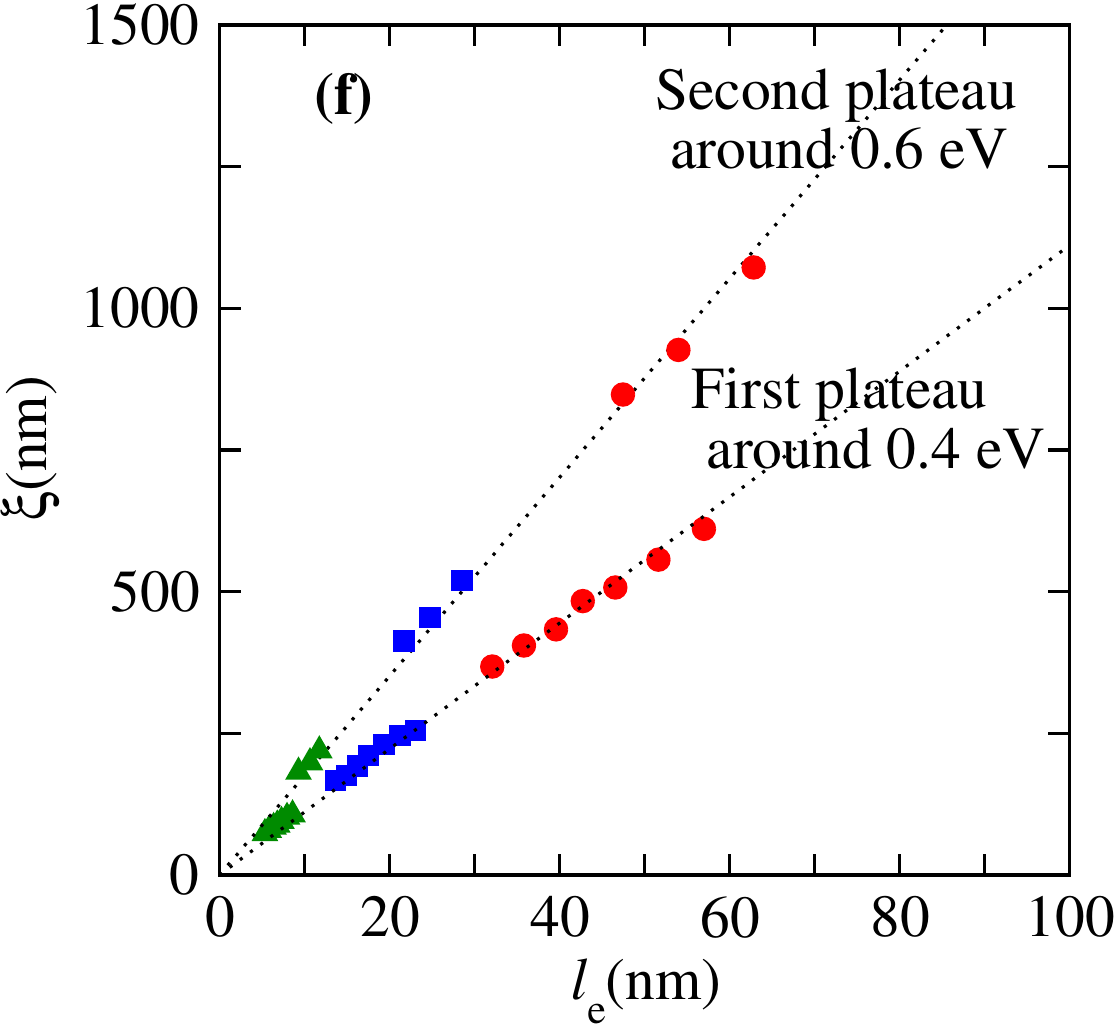}
\caption{Low concentration of resonant adsorbates: 0.1\% (red/filled circle), 0.2\% (blue/filled square) and 0.4\% (green/filled triangle). (a) Density of states versus energy, (b) Elastic mean-free path $ l_{e} $ versus energy, (c) Microscopic conductivity $ \sigma_{M} $ versus energy, (d) Electronic conductivity $ \sigma $ versus inelastic mean-free path $ L_{i} $ for 0.1\% of resonant adsorbates, (e) Electronic conductivity $ \sigma $ versus inelastic mean-free path $ L_{i} $ at $E=0.4$\,eV, and (f) Localization length $ \xi $ versus elastic mean-free path $ l_{e} $ (Thouless relation).}
\label{fig_low_resonant}     
\end{figure}

%%%%%%%%%%%%%%%%
\begin{figure}
    \centering
    \includegraphics[height=5cm,width=5cm]{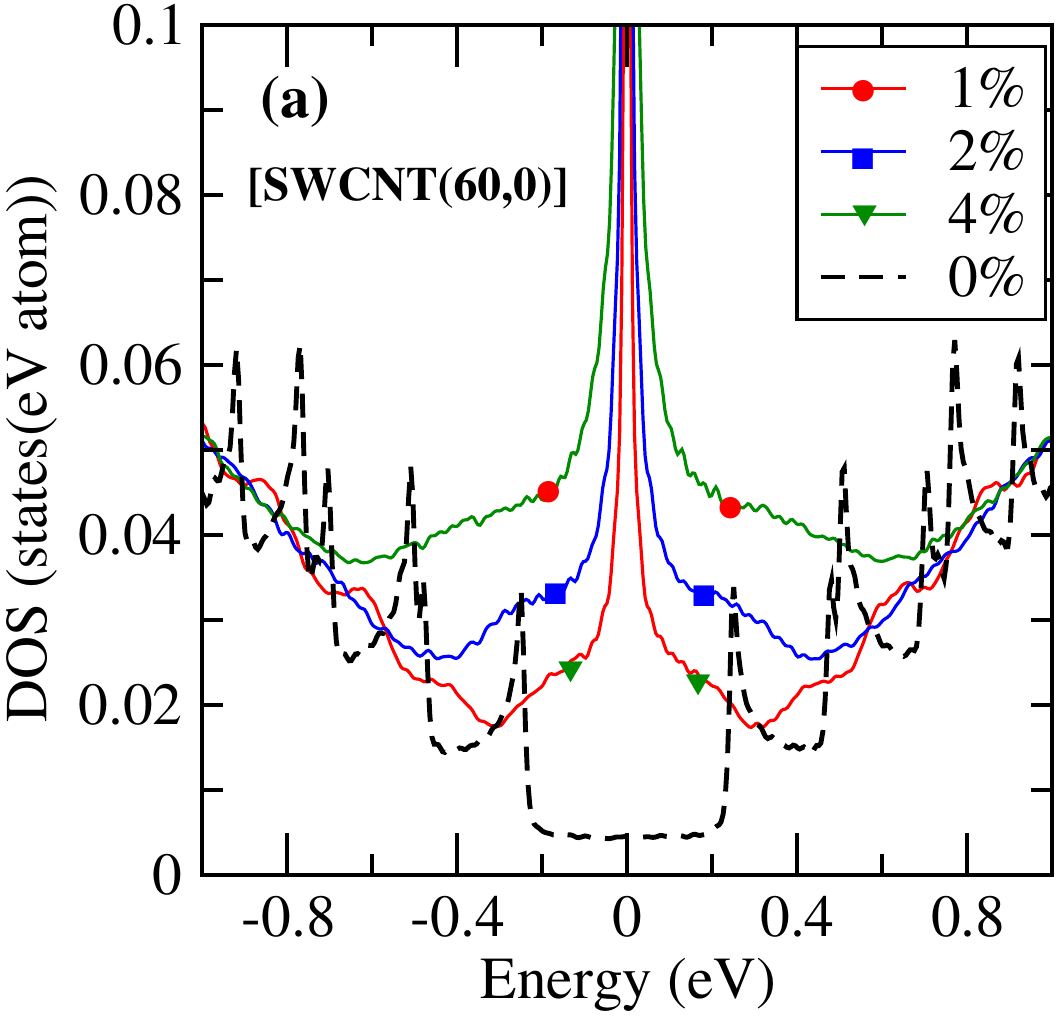}
    \includegraphics[height=5cm,width=5cm]{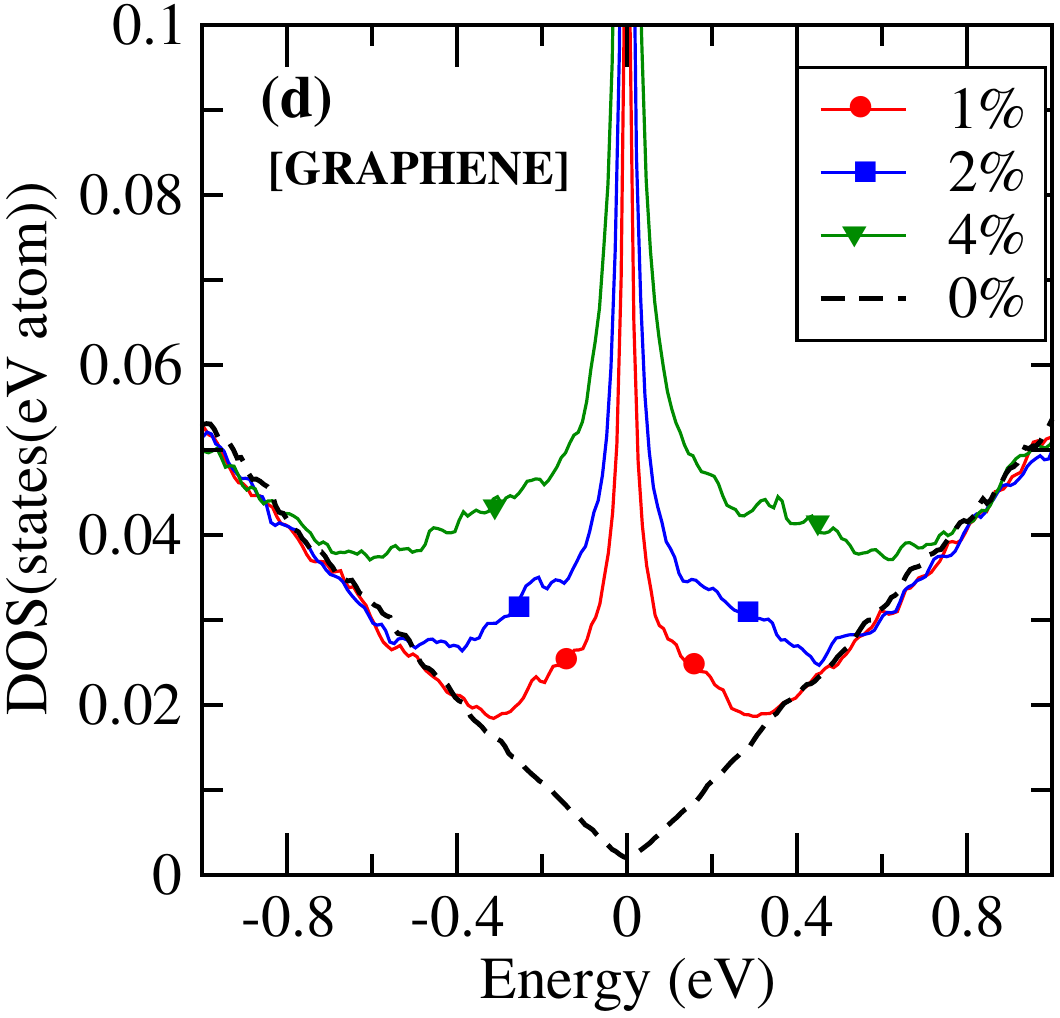} \\
    \includegraphics[height=5cm,width=5cm]{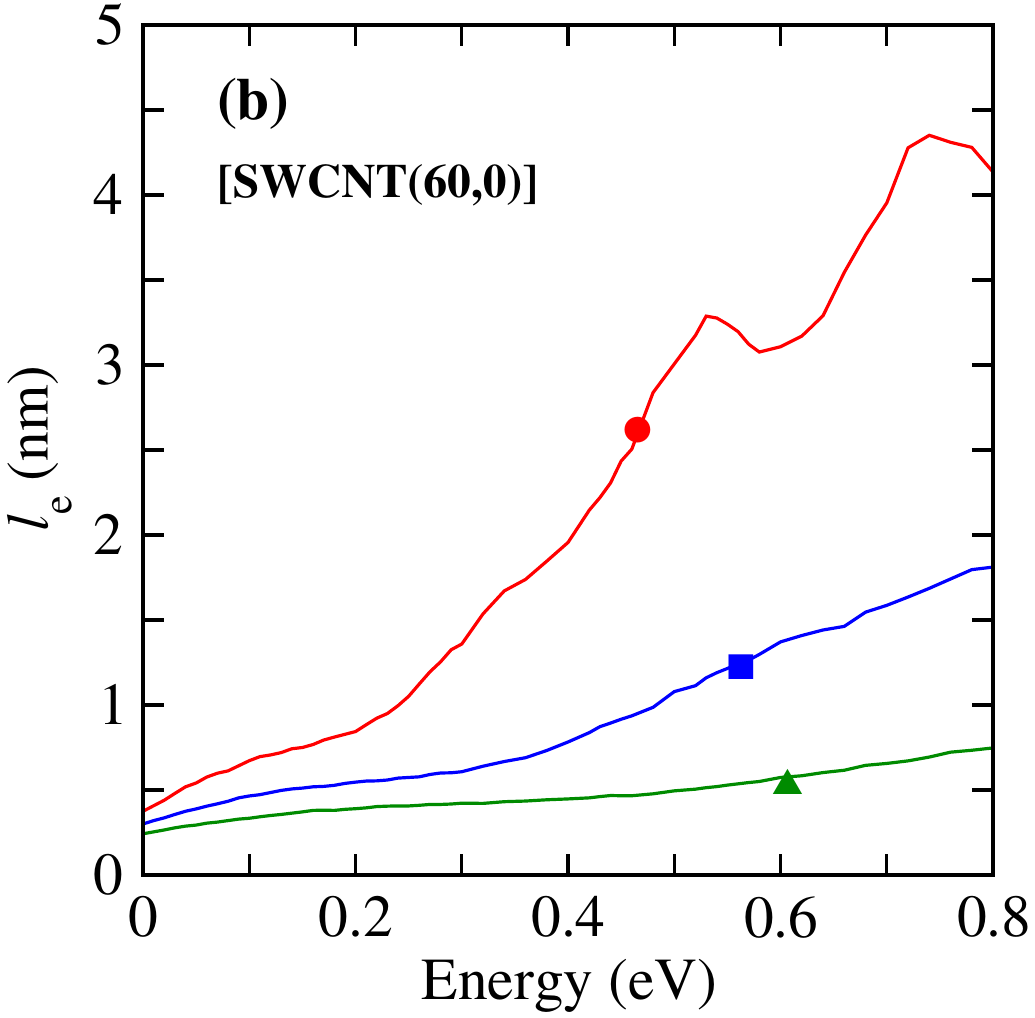}
    \includegraphics[height=5cm,width=5cm]{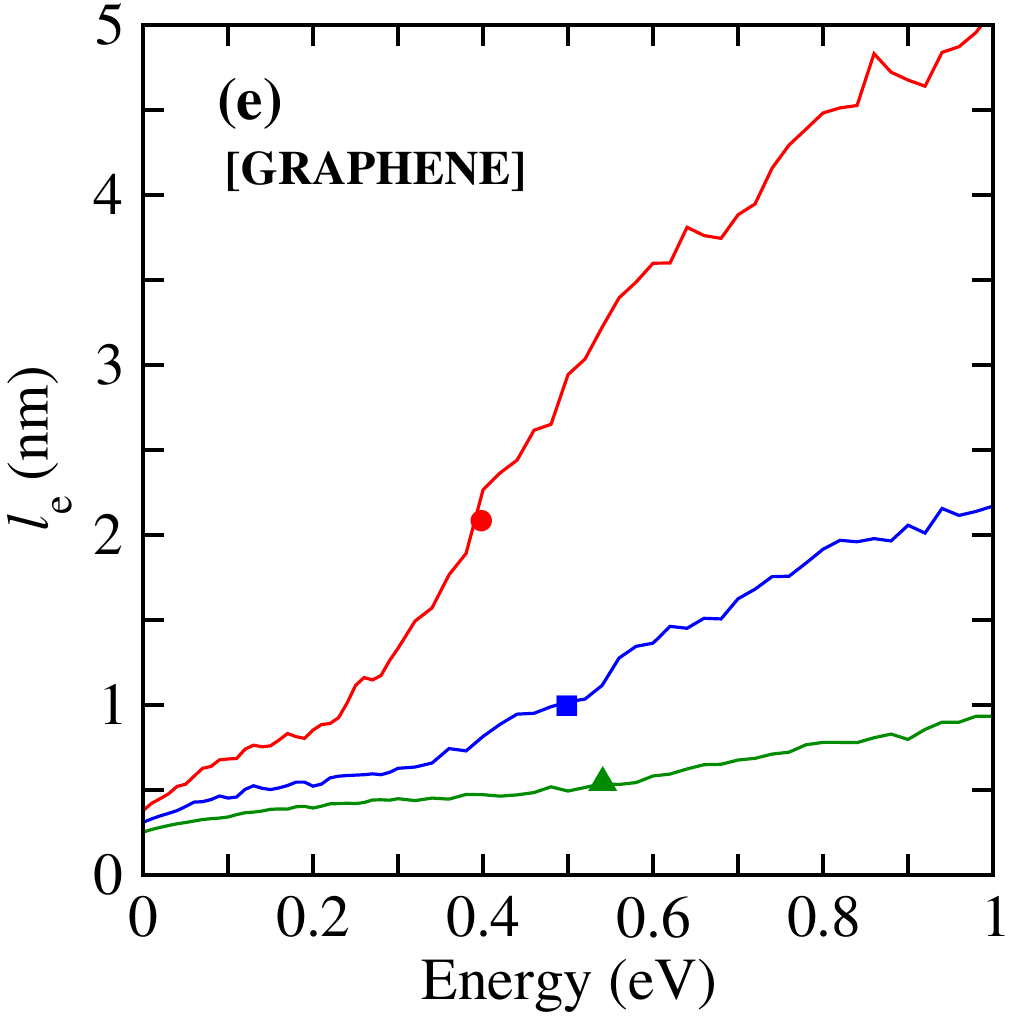}\\
    \includegraphics[height=5cm,width=5cm]{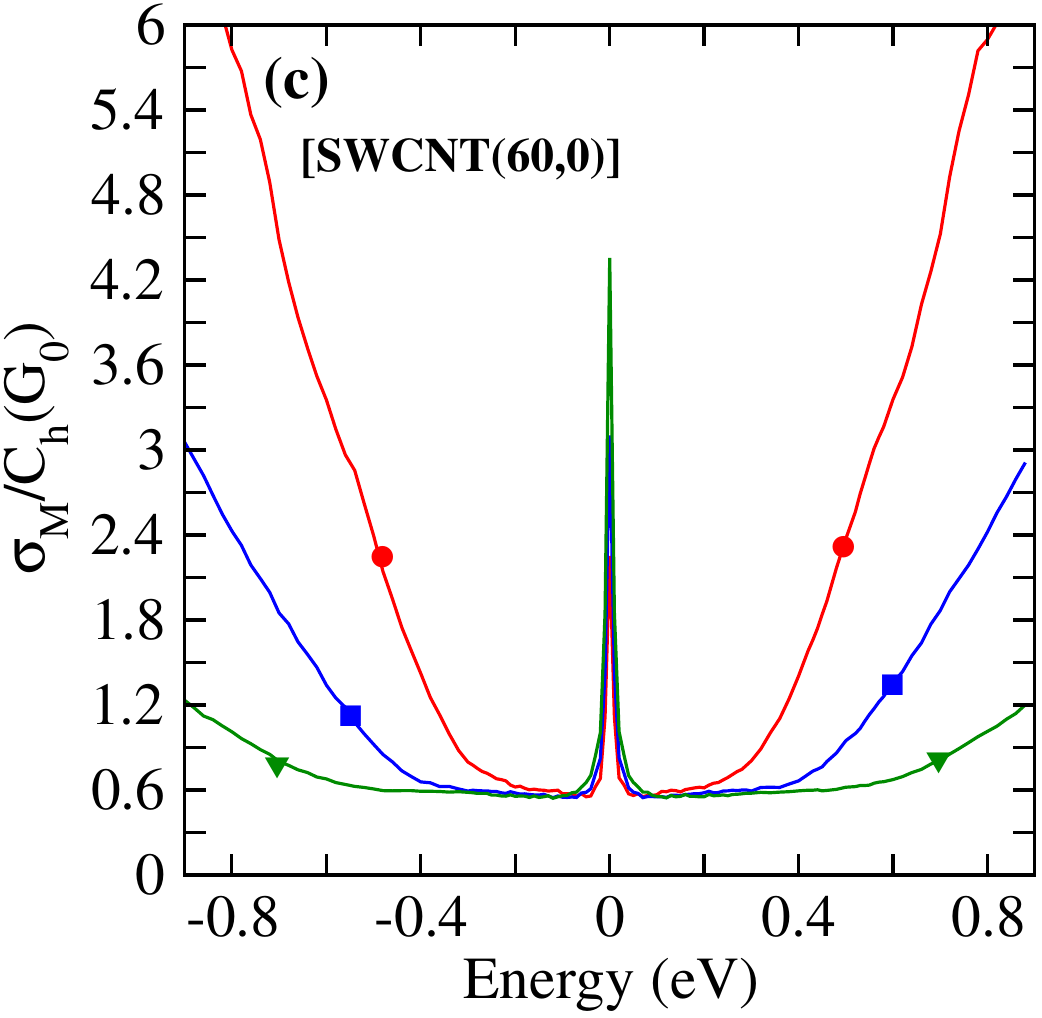}
    \includegraphics[height=5cm,width=5cm]{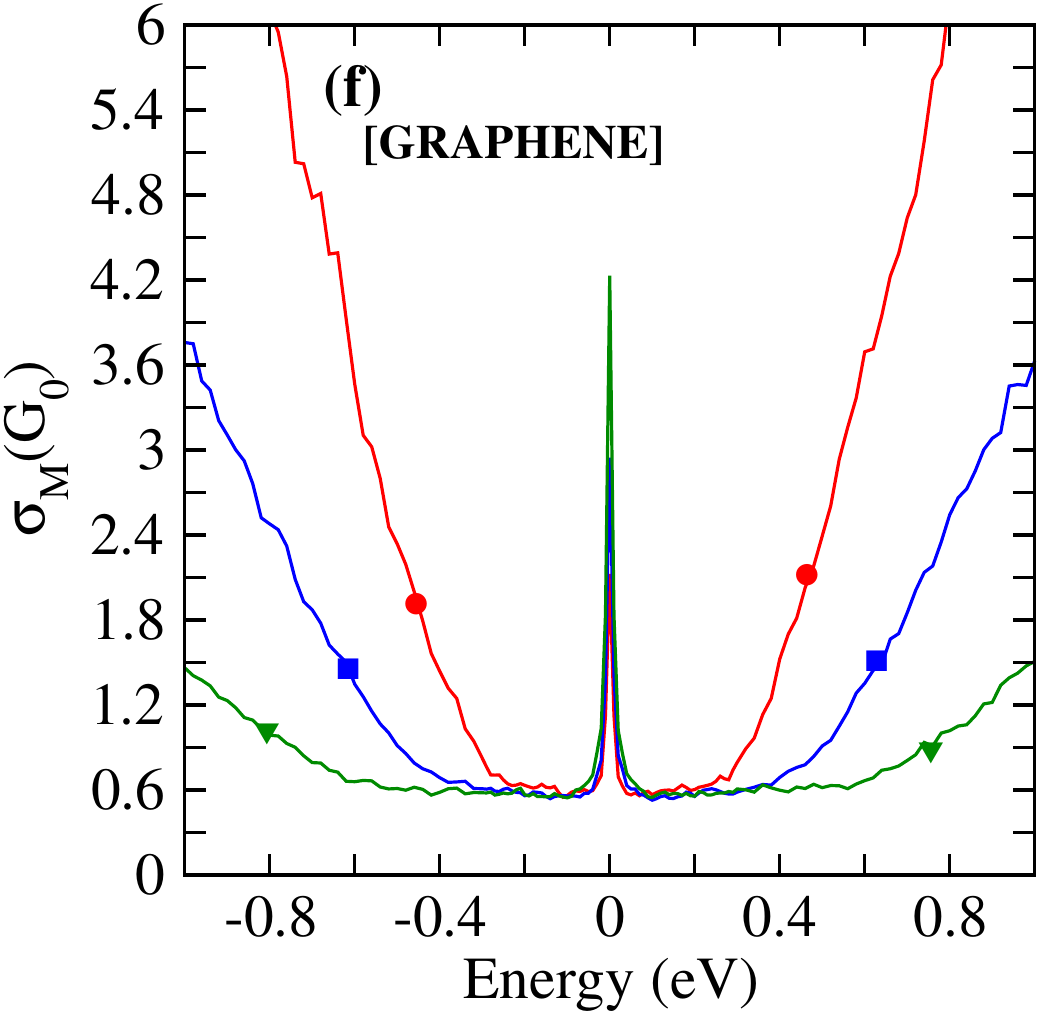}

\caption{A comparative representation of electronic transport properties of  SWCNT (60,0) and graphene at high concentration of resonant adsorbates: 1\% (red/filled circle), 2\% (blue/filled square) and 4\% (green/filled triangle). [SWCNT(60,0)] (a) Density of states, (b) Elastic mean-free path $ l_{e} $, and (c) Microscopic conductivity per unit of surface, $ \sigma_{M}/C_{h} $, versus energy. [GRAPHENE] (d) Density of states, (e) Elastic mean-free path $ l_{e} $, and (f) Microscopic conductivity $ \sigma_{M} $, versus energy. }     
\label{fig_high_resonant}
\end{figure}

We now investigate the electronic behavior of the metallic SWCNT $\left(60,0\right)$
in presence of resonant  adsorbates. We consider first the low concentration cases 
(0.1\%, 0.2\% and 0.4\%) (Fig. \ref{fig_low_resonant}) and then higher concentrations cases (1\%, 2\% and 4\%) (Fig. \ref{fig_high_resonant}).

In Fig. \ref{fig_low_resonant}-a, the density of states versus energy for resonant adsorbates shows more dramatical effects with a pic of resonant states around the Fermi's energy, a shift in the plateau of the DOS for larger energy and the positions and height of the VHS are greatly affected. 

For resonants adsorbates as shown in Fig. \ref{fig_low_resonant}-b, the mean-free path increases as a function of energy within a plateau of energy and by crossing a VHS, since the probability of scattering is much higher for low energies close to the resonant states where $l_{e}$ is at its minimal value. $l_{e}$ is shown to be always larger than the circumference of the nanotube which confirms that the electronic behavior of SWCNT $\left(60,0\right)$ is one dimensional.

The microscopic conductivity $\sigma_{M}$ expressed in unit of quantum conductance $G_{0}={2e^{2}}/{h}$ versus energy is shown in 
Fig. \ref{fig_low_resonant}-c for low concentration of resonant adsorbates. The
increase of concentration of adsorbates decreases the conductivity
$\sigma_{M}$. The evolution of $\sigma_{M}$ as a function of energy, follows the general behavior of $l_{e}$ as a function of energy. $\sigma_{M}$ is a semi-classical quantity computed in the diffusive regime when $L_{i}$ and $l_{e}$ are comparable. Thus, it does not take into account the quantum effects (localization effects). The right column of Fig. \ref{fig_low_resonant} shows the occurence of localization effects similar to those shown for the non-resonant states. This behavior corresponds to the laws of one-dimensional conductors as given by equations (\ref{eq3}) and (\ref{Th}).

Figure \ref{fig_high_resonant}-a, shows that at high concentration of resonant adsorbates (1\%, 2\% and 4\%), the VHS disappear from the density of states around the Fermi's energy, because the effects of electronic scattering is so dramatic to the point that the quantification of the $\vec{k}$ vector in the direction of $\vec{C}_{h}$ is broken. The new shape of the DOS is actually close to the DOS of graphene shown in Fig. \ref{fig_high_resonant}-d. 

We can see in  Fig.  \ref{fig_high_resonant}-b that the elastic mean-free path is much shorter than the circumference $C_{h}$, for all energies. As a result, we can expect a transition from one-dimensional towards two-dimensional electronic behavior in a SWCNT induced by increasing the concentration of adsorbates. Indeed if the characteristic lengths are shorter than its circumference the nanotube should behave like a sheet of graphene.

This dimensionality crossover should be identified by investigating the effects on the microscopic conductivity and the scale dependent conductivity. Since the beginning, the electrical conductivity $\sigma$ is expressed in units of length for one-dimensional system $\sigma^{1D}$, we redefine this quantity for a two-dimensional system by dividing it by $C_{h}$, so $\sigma^{2D}=\frac{\sigma^{1D}}{C_{h}}$ is expressed
in units of surface. 

As shown in Fig. \ref{fig_high_resonant}-c  the high concentration of resonant adsorbates leads to a plateau of minimum conductivity around  $0.6\,G_{0}$, similar to the reported results on the 2D graphene (see Fig. \ref{fig_high_resonant}-f) in other studies \cite{trambly_graphene_2013}.
This value of minimum conductivity is universal as long the diameter of zigzag SWCNT is large enough and the concentration of resonants adsorbates is high enough to actually create a dimensionality crossover in the electronic behavior.

%%%%%%%%%%%%%%%%
\begin{figure}
    \centering
    \includegraphics[width=12cm]{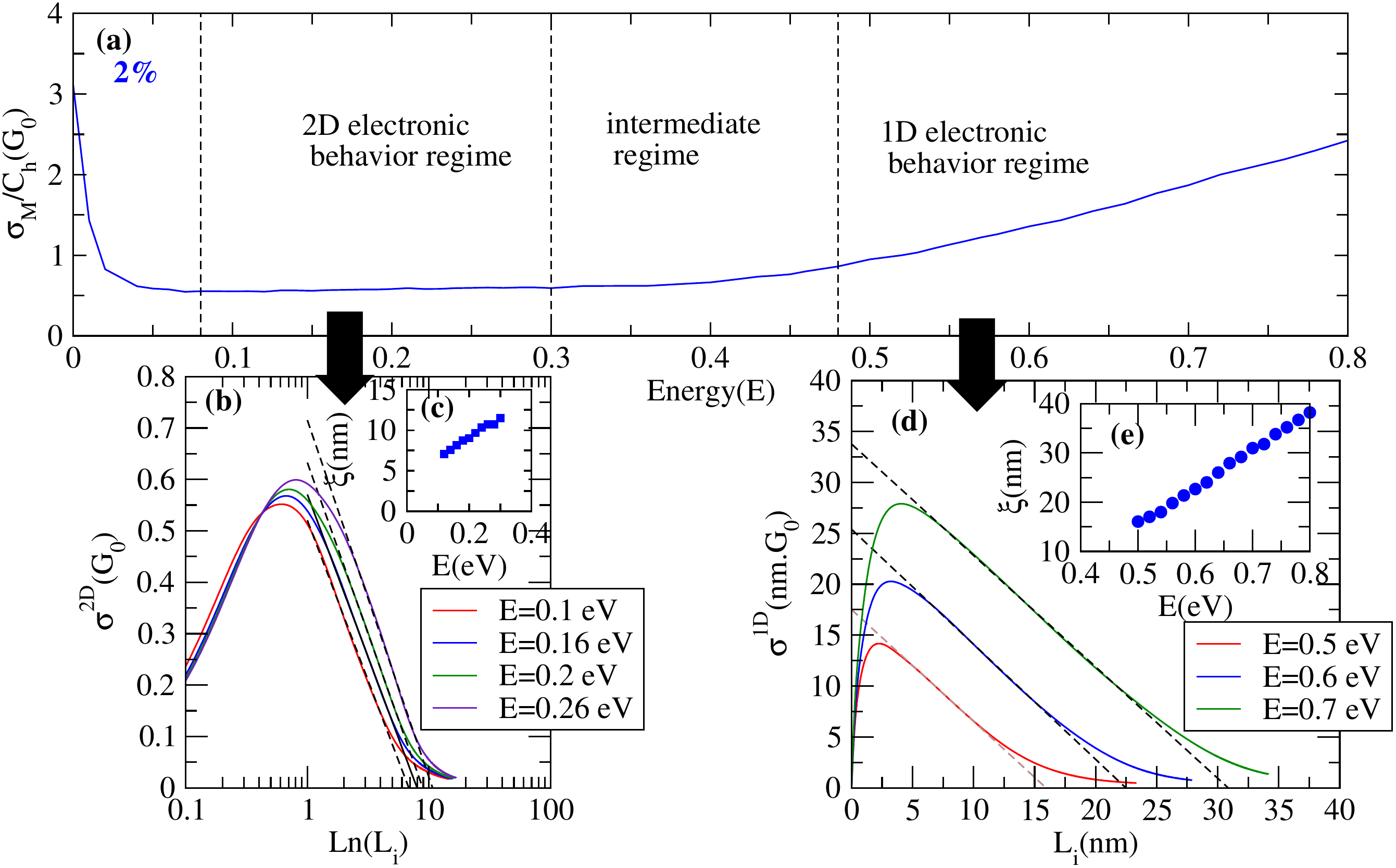} 
    
\caption{For SWCNT (60,0) with 2\% of resonant defects: (a) Microscopic conductivity per unit of surface $ \sigma_{M}/C_{h} $ versus energy, (b) Electronic conductivity versus inelastic mean-free path in log scale, (c) Localization length versus energy shows 
{%\textcolor{red}{
$ \xi< C_{h} $, 
}
(d) Electronic conductivity versus inelastic mean-free path, and (e) Localization length versus energy shows $ \xi>C_{h} $. }
\label{fig_1D-2D}     
\end{figure}

At 2\% of resonant adsorbates, Fig. \ref{fig_1D-2D}
shows clearly  the transition of the electronic behavior in the microscopic conductivity. For low energy in Fig.  \ref{fig_1D-2D}-b the 2D scale dependent conductivity: $\sigma_{2D}\left(L_{i}\right)=\sigma_{0}-\beta G_{0}\ln\left(\frac{L_{i}}{l_{e}}\right)$,
fits the decreasing behavior of the electronic conductivity towards
the localization regime in the logarithmic scale for large $L_{i}$.
The slope in log scale according to $\sigma_{2D}\left(L_{i}\right)$
shows a new universal slope, $\beta={1}/{\pi}$, similar to what
we observe in 2D graphene \cite{trambly_graphene_2013}, which confirms the 2D electronic behavior in presence of high concentration of resonant adsorbates. Based on our estimation of localization length shown in Fig. \ref{fig_1D-2D}-c, this 2D electronic behavior appears when the localization length is shorter than the circumference $\xi<C_{h}$. 
For high energy as shown in Fig. \ref{fig_1D-2D}-d,
the 1D scale dependent conductivity confirmes the 1D electronic behavior when $\xi>C_{h}$. Therefore the 1D/2D crossover is governed by the ratio between the localisation length $\xi$ and the circumference $C_{h}$. Thus we can define as a function of energy within the microscopic conductivity three regimes $\left(1\right)$a 2D regime when $\xi<C_{h}$, $\left(2\right)$ An intermediate regime when $\xi\approx C_{h}$and $\left(3\right)$ A 1D regime when $\xi>C_{h}$.

\section{Discussion}
\label{sec_discussion}

We now discuss the case of non-resonant and resonant adsorbates by combining the results on the (60,0) nanotube and those briefly presented in the  supplementary material on (30,0) and (90,0) nanotubes. We focus particularly on the localization length which is a measure of the importance of localization effects and their impact on transport. Strictly at low temperature the conductance of a nanotube becomes exponentially small  if the length of the nanotube is larger than the localization length. Equivalently if this length is smaller than the localization length then the nanotube can conduct. Yet at finite temperature the inelastic scattering can destroy the localization effect on a length $L_{i}$. Therefore if $\tau_{i}$ is smaller than the time needed for localization $\tau_{L}$ the nanotube can conduct. We have discussed this in the supplementary material and find that the time needed for localization $\tau_{L}$  varies between $10^{-12}$\,s   and a few $10^{-11}$\,s.

For non-resonant scatterers we find that in all cases the Anderson localization is well described by the standard  1D transport theory in a low concentration limit. This implies that the elastic mean-free path obeys the Fermi golden rule and varies inversely proportionally to the adsorbate concentrations. In addition the relation between the elastic mean-free path and the localization length obeys the Thouless relation. Depending on the Fermi energy and the adsorbate concentrations the localization length varies between $1200$\,nm and $30$\,nm. These values are similar for all three nanotubes. Yet a noticeable difference is that  for smaller diameters the energy range where there are only two channels is more important. In this two channels zone the localization length tends to be larger than at higher energies. Therefore nanotubes with non-resonant adsorbates will present better transport when the Fermi energy is closer to the CNP.

For the resonants adsorbates the situation is opposite. Indeed the scattering matrix $T(E)$  presents a strong resonance close to the charge neutrality point and this leads to a short mean-free path and a short localization length  at energies close to the charge neutrality point. Therefore in the case of resonant adsorbates the transport properties are more deteriorated close to the CNP and better transport properties will be obtained if the Fermi energy is sufficiently far from the CNP. At low concentrations the behaviour is also well described by the 1D quantum transport theory. except very close to the neutrality point where even the density of states is modified by the adsorbates. However at higher concentrations and for sufficiently large diameter  carbon nanotubes can behave like a 2D system and not like a 1D conductor. This happens when the elastic mean-free path and the localization length become smaller than the circumference of the nanotube. For large nanotubes with diameter of 4-5 nanometers this regime can be reached with resonant scatterers at concentrations of defects of the order of one percent and for energies of a fraction of an electron Volt away from the charge neutrality level ($E<0.3$\,eV at  $2\%$). This 2D behaviour is comparable with the electronic behaviour in graphene, in which $\sigma\left(L_{i}\right)$ decreases for large $L_{i}$ accordingly to the 2D scale dependent conductivity and leading to the creation of minimum microscopic conductivity around the Dirac energy. Yet at higher energies ($E>0.5$\,eV at  $2\%$) the scattering by resonant states is less efficient and one recovers a typical 1D behaviour.  Depending on energy and concentration the localization length varies between $1000$\,nm and $10$\,nm. These values are about the same in all three nanotubes. 

Finally we discuss the possibility of a transport mechanism that is based on hopping. This type of transport is well documented in nanotubes and bundles or two dimensional arrays of nanotubes. At the very low temperatures of a few Kelvin, the electron-electron interaction can play a strong role but at higher temperatures the  thermally activated hopping between localized states is usually due to  electron-phonon coupling \cite{roche_charge_2007, xu_phonon_2013}. We shortly discuss this regime now.

%%%%%%%%%%%%%%%%
\begin{figure}
    \centering
    \includegraphics[height=5cm,width=5cm]{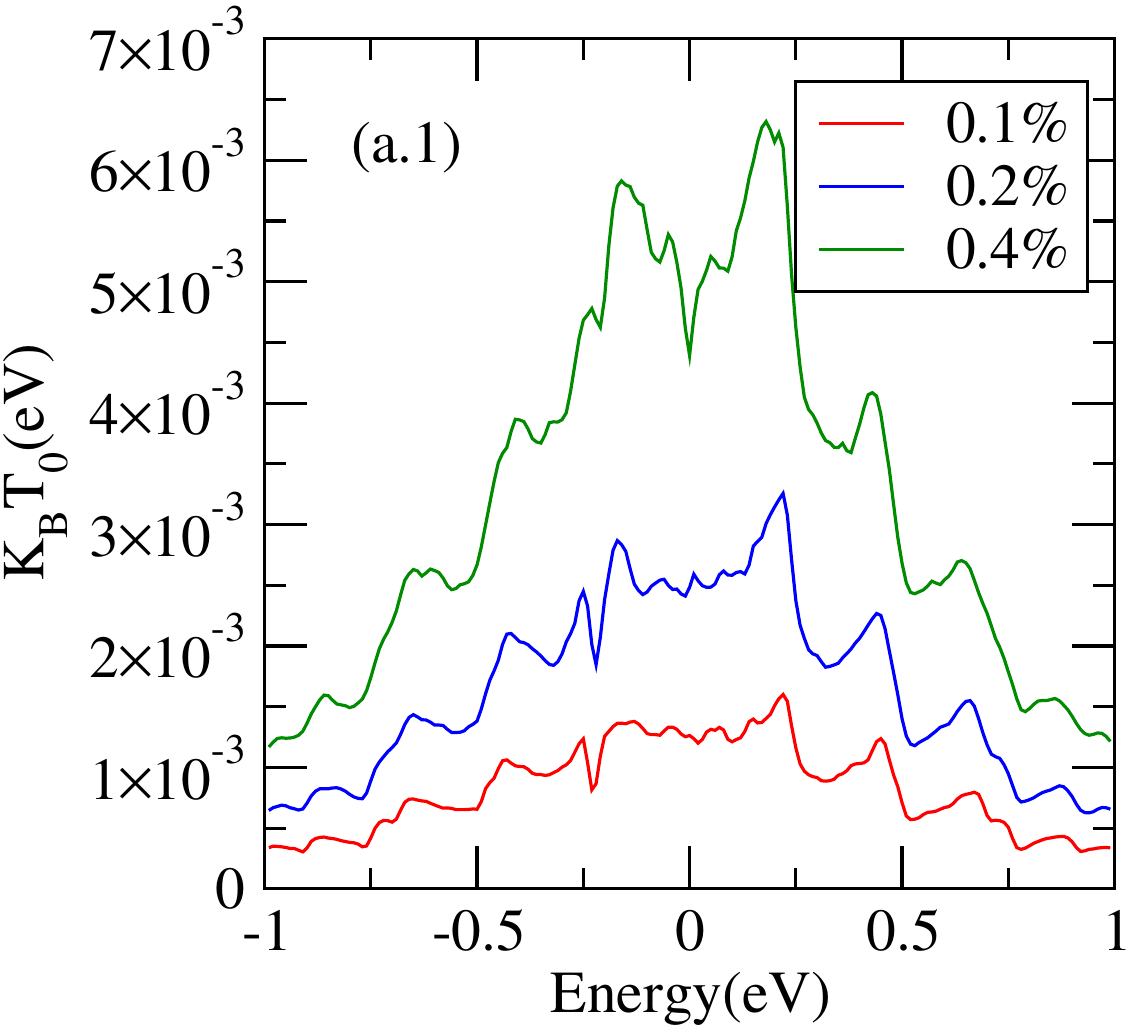} 
    \includegraphics[height=5cm,width=5cm]{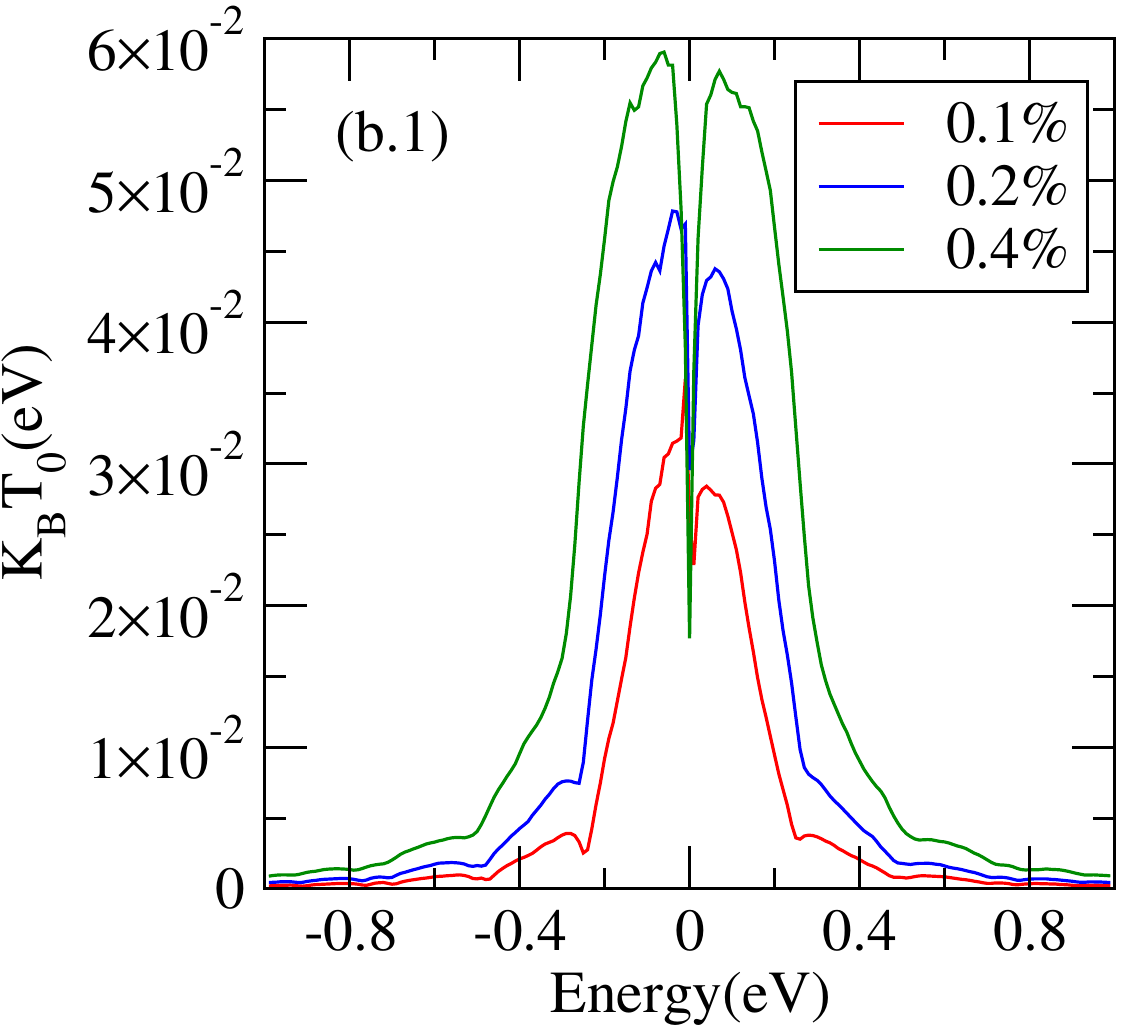}\\
    \includegraphics[height=5cm,width=5cm]{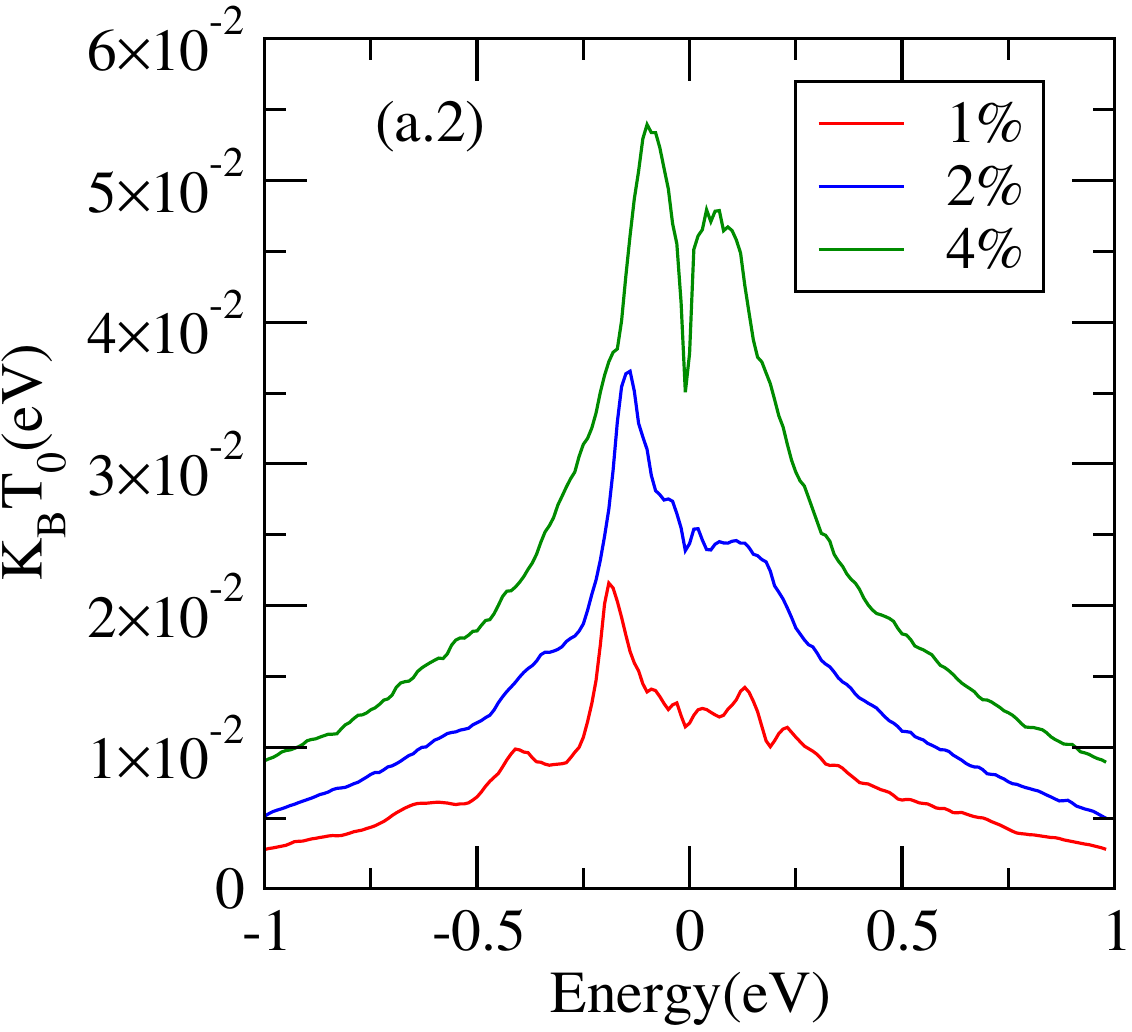}
    \includegraphics[height=5cm,width=5cm]{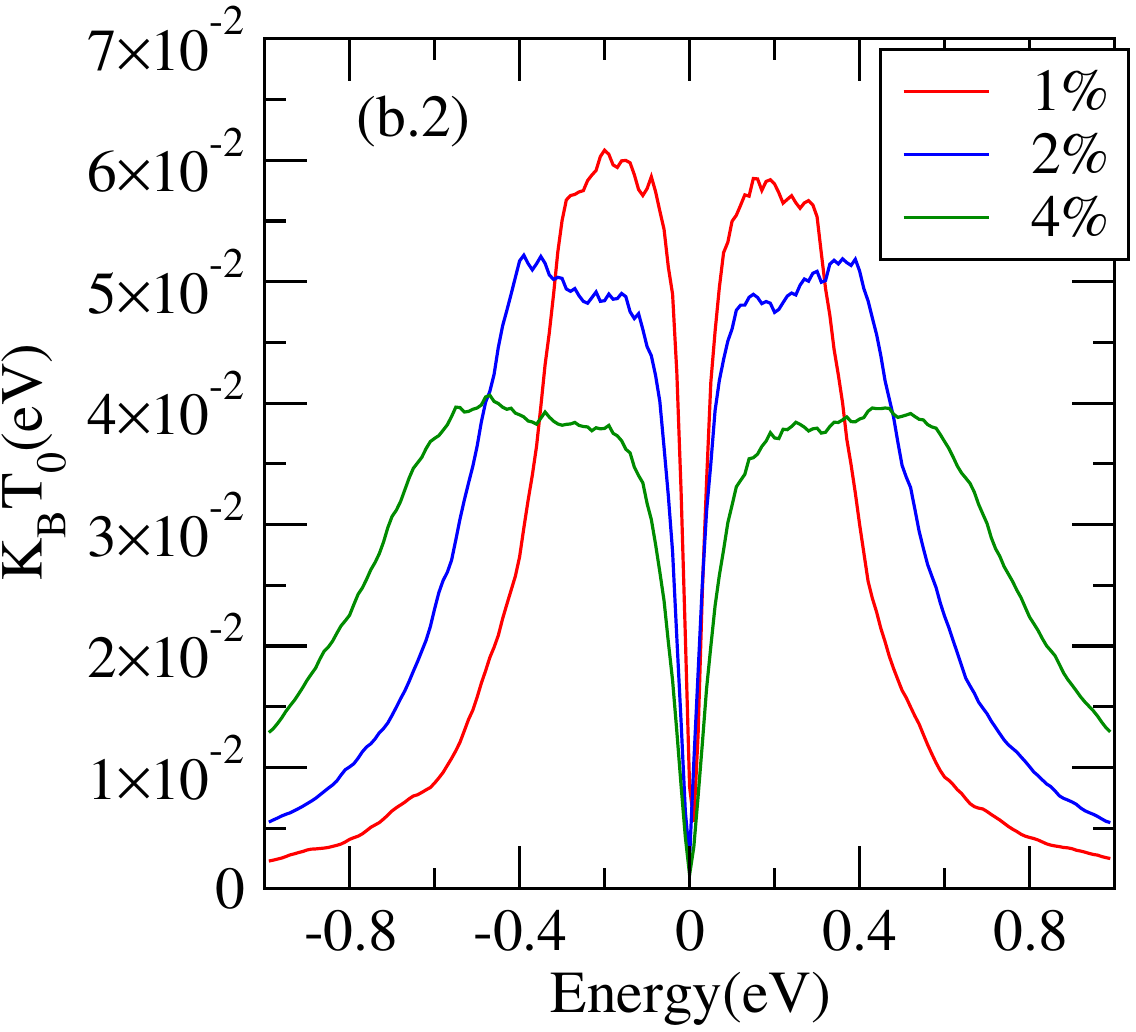}

\caption{Temperature  $k_B T_0 $ as function of energy $E$ for low and high concentrations. Results in the left column are for non-resonant defects and  in the right column for resonant defects. Room temperature $T_R$ corresponds to $k_B T_R \simeq 2.5\times10^{-2}$\,eV. }
\label{fig_KT}     
\end{figure}
%%%%%%%%%%%%%%%%

In the nearest neighbor hopping regime the transport is thermally activated and in the  variable range hopping regime (which occurs  at lower temperature)  the conductivity $ \sigma $ is typically proportional to $ \propto \exp(-(T_0/T)^{1/2})$ \cite{yu_variable_2001, wang_hopping_2007, pipinys_VRH_2012}. This regime of variable range hopping appears typically at temperature below $T_0 = 4/(n(E)\xi (E))$ where $\xi (E)$ is the localization length and $n(E)$ the density of states per unit length and per spin. Fig.  \ref{fig_KT} shows the variation of this temperature $T_0 $ as a function of concentration and energy $E$ for resonant and non-resonant defects. These results show that transport through the Variable Range Hopping  transport could occur even at room temperature in systems with sufficiently large defects concentration provided that the Fermi energy is sufficiently close to the charge neutrality point. Note that for resonant states the $T_0 (E)$ decreases in the immediate vicinity of the CNP due to the higher density of states caused by the resonant states.
Around the CNP, $T_0 (E)$ increases as function of concentration of resonant and non-resonant adsorbates, however beyond 1\% of resonant adsorbates (at 2\% and 4\%) as shown in Fig.  \ref{fig_KT}-b.2, $T_0 (E)$ decreases which coincide with the appearance of the strong 2D electronic behavior previously discussed.

\section{Conclusion}

We have presented a study of  the quantum localization in  single walled carbon nanotubes (SWCNT), with resonants and non-resonants adsorbates that represent two extreme types of covalently functionalized groups with moderate and strong scattering properties. In the present study the concentration of these adsorbates varies between 0.1\%,  and 4\% and we consider nanotubes  of average and large circumferences (up to $ 15$\,nm). Our study shows that the type of adsorbates, the circumference of the nanotube and the Fermi energy are determinant for the localization phenomenon. Depending on these parameters the localization length can become small of the order of a few $ 10$\,nm. The localization can even be smaller than the circumference of the nanotube in the case where there are resonant scatterers, a Fermi energy close to the CNP and a large nanotube. In this special case the localization effect follows the 2D laws. Quite generally we find that non-resonant adsorbates have better transport properties for Fermi energy close to the CNP where it is the opposite with resonant adsorbates. These systems could also present a transport through nearest neighbor hopping or even through Variable Range Hopping even at room temperature provided that the Fermi energy is sufficiently close from the Charge Neutrality Point and also for sufficiently high concentration of adsorbates. 

We believe that the present study can help to understand the localization effect in a given device and to determine in which regime it is. Therefore this work should be useful for obtaining efficient nano-sensors based on covalent functionalisation of carbon nanotubes.

\section*{Acknowledgments}
The authors wish to thank 
Petrutza Anghel Vasilescu and Ahmed Missaoui
for fruitful discussions.
The numerical calculations have been performed 
at  Institut N\'eel, Grenoble,
and at the Centre de Calculs (CDC),
Universit\'e de Cergy-Pontoise.
We thank Y. Costes, CDC, for computing assistance. 
JJK thanks the Institute for advanced studies, Universit\'e of Cergy-Pontoise, for financial support.

\section*{References}

\bibliography{references}
\bibliographystyle{unsrt}

%\includepdf[pages={1}]{myfile.pdf}


\section*{Supplementary material}

*Supplementary Material for*

# Quantum localization and electronic transport in covalently functionalized carbon nanotubes

Ghassen Jemaï ,[1]* Jouda Jama Khabthani,[1] Guy Trambly de Laissardière[2]

and Didier Mayou[3]

[1] Laboratoire de Physique de la Matière Condensée, Département de Physique, Faculté des Sciences de Tunis, Université Tunis El Manar, Campus Universitaire, 1060 Tunis, Tunisia

* ghassen.jemai@fst.utm.tn

[2] Laboratoire de Physique Théorique et Modélisation, CNRS and Université de Cergy-Pontoise, 95302 Cergy-Pontoise, France

[3] Université Grenoble Alpes, CNRS, Inst NÉEL, 38042 Grenoble, France

(Dated: August 7, 2019)

In Sec I of this Supplemental Materials, we briefly present the computational method to calculate electrical conductivity in presence of static defects, and within the relaxation time approximation (RTA). In Sec. II, results for single walled carbon nanotubes (SWCNT) (30,0) are presented. And the 1D/2D transition for SWCNTs (30,0) and (90,0) is discussed in Sec III.

## I. NUMERICAL METHOD

In recent studies on low dimensional carbon systems such as graphene monolayer and bilayer, it has been reported that the real space Kubo-Greenwood formalism computed by recursion in real space and polynomial expansion method [1–5], allows us to compute efficiently quantum transport in a large system with up to few $10^7$ atoms which is useful to study systems with electronic characteristic length of the order of a few hundred to few thousand of nanometers [1–9]. We compute the Kubo-Greenwood formula using the Einstein relation between the electronic conductivity and the quantum diffusion

$$\sigma(E_F) = e^2 n(E_F) D(E_F), \qquad (1)$$

1

where $e$ is the electron charge, $n$ is the density of states, $E_F$ is the Fermi's energy and $D$ is the quantum diffusivity given by

$$D(E_F) = \frac{1}{2} \lim_{t \to \infty} \frac{d}{dt} X^2(E_F, t). \tag{2}$$

$X^2(E, t)$ is the average square spreading of the electron wave packet at energy $E$ and after a propagation time $t$,

$$X^2(E, t) = \frac{\text{Tr}\left\{ \left[ \tilde{X}(t) - \tilde{X}(0) \right]^2 \delta(E - H) \right\}}{\text{Tr}\left[ \delta(E - H) \right]}, \tag{3}$$

where $\tilde{X}(t)$ is the position operator for the electron and $\delta(E - H)$ is the projection operator at energy $E$. The quantum diffusion is also related to the velocity correlation function $C(E, t)$ by the relation

$$\frac{d}{dt}\left( X^2(E, t) \right) = \int_0^t C(E, t') dt', \tag{4}$$

where the velocity correlation function is given by

$$C(E, t) = \left\langle \tilde{V}_x(t) \tilde{V}_0(t) + \tilde{V}_x(0) \tilde{V}_x(t) \right\rangle_E = \Re \left\langle \tilde{V}_x(t) \tilde{V}_0(t) \right\rangle_E. \tag{5}$$

Here $\tilde{V}_x(t)$ is the velocity operator in the Heisenberg representation along the $x$-direction defined as

$$\tilde{V}_x = \frac{1}{i\hbar} \left[ \tilde{X}, H \right]. \tag{6}$$

The formalism we use here treats exactly the elastic process, where the elastic mean-free path is $l_e = D_{max}/V_G$, where $D_{max}$ is the maximum value of diffusivity and $V_G$ the electron velocity in graphene. The inelastic scattering events destroy the phase coherence after a characteristic time scale $\tau_i$ associated to a length scale called the inelastic mean-free path $L_i(E, \tau_i)$. The inelastic phenomena is taken into account in a phenomenological way using the relaxation time approximation (RTA). Within the RTA we assume that the velocity correlation function $C_i(E, t)$ (with inelastic scattering) is given by [6, 7]

$$C_i(E, t) \simeq C(E, t) \, e^{\frac{-|t|}{\tau_i}}. \tag{7}$$

Thus the inelastic mean-free path $L_i$ is computed based on the average square spreading is written as

$$L_i^2(E, \tau_i) = \frac{1}{2\tau_i} \int_0^\infty \Delta X^2(E, t) e^{-t/\tau_i} dt, \tag{8}$$

with the diffusivity is given by

$$D(E, \tau_i) = \frac{L_i^2(E, \tau_i)}{\tau_i}.$$

Finally, the expression of electrical conductivity becomes

$$\sigma(E_F, \tau_i) = e^2 n(E_F) D(E_F, \tau_i).$$

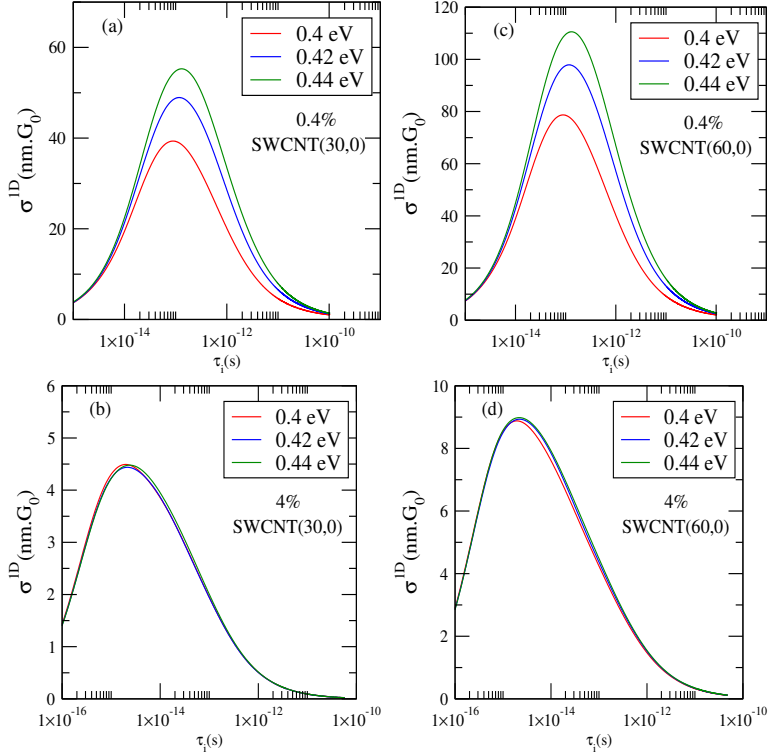

FIG. 1. Electronic conductivity versus inelastic time: (a) SWCNT (30,0) with 0.4% of resonant adsorbates, (b) SWCNT (30,0) with 4% of resonant adsorbates, (c) SWCNT (60,0) with 0.4% of resonant adsorbates, and (d) SWCNT (60,0) with 4% of resonant adsorbates.

In a low dimensional system (quasi 1D) various quantum phenomena take place such as coherent backscattering and the destructive quantum interference of different possible paths of the scattered electron wave packet, which leads to Anderson localization that occurs on a length scale shorter than the phase coherence length scale $\xi < L_\phi$ [10, 11]. Without any inelastic process this means that the sample become insulator. But this is not the case here since the effect of the inelastic process leads to a loss of phase coherence on a length scale $L_i < \xi$. The variation of the conductivity $\sigma(E_F, \tau_i)$ with $\tau_i$ or with $L_i(E, \tau_i)$ contains essential information about the localization effects.

3

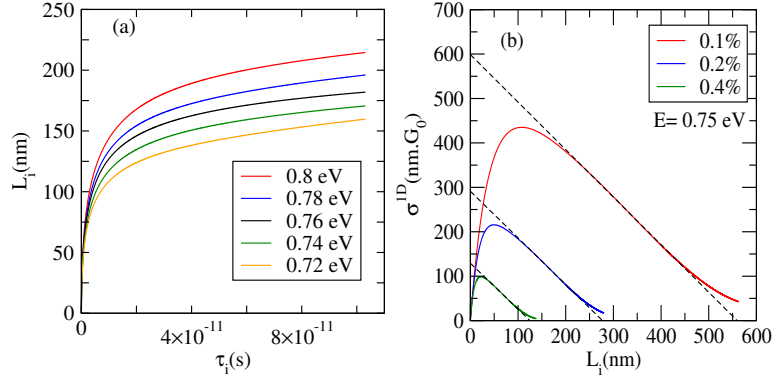

FIG. 2. SWCNT (30,0) with resonant adsorbates: (a) inelastic mean-free path versus inelastic time with 0.2% of resonant adsorbates around the energy $E = 0.7\,\text{eV}$, and (b) electronic conductivity versus inelastic mean-free path.

In figure 1 we show the electronic conductivity $\sigma(E, \tau_i)$ versus inelastic time scattering $\tau_i$ for various adsorbate concentrations and different carbon nanotubes (SWCNT (30,0) and (60,0)). At small $\tau_i$, $\sigma(E, \tau_i)$ increases as function of inelastic time and the electronic transport is ballistic until reaching a maximum which is the microscopic conductivity. For large $\tau_i$ the quantum effects of localization force $\sigma(E, \tau_i)$ to decreases rapidely in the localization regime. The time needed to reach the localization $\tau_L$ varies between few $10^{-12}\,\text{s}$ to a few $10^{-11}\,\text{s}$. Note also that the localization length is in principle numerically estimated using the relation $\xi(E) = \frac{1}{\pi} \lim_{t \to \infty} \sqrt{X^2(E, t)}$ [12] (where $X^2(E, t)$ is equivalent to $L_i^2(E, t)$ within the RTA). However the method is too expensive numerically and it takes a lot of time to reach localization that is the saturation of $X(E, t)$ ($L_i(E, t)$) as shown in figure 2-a. Instead by computing $\sigma(L_i)$ the electronic conductivity as a function of inelastic mean-free path $L_i(E, \tau_i)$ one obtains the relation of the scale dependent conductivity for large $L_i$. The extrapolation of this relation as shown in figure 2-b gives us a precise estimation of the localization length.

## II. SWCNT (30,0) WITH NON-RESONANT AND RESONANT ADSORBATE

SWCNT (30,0) has a diameter $d = 2.37\,\text{nm}$ and a circumference $C_h = 7.47\,\text{nm}$ which makes it half the size of the SWCNT (60,0) ($C_h = 14.94\,\text{nm}$). Here we present briefly some results for SWCNT (30,0) with resonant and non-resonant adsorbates, comparing

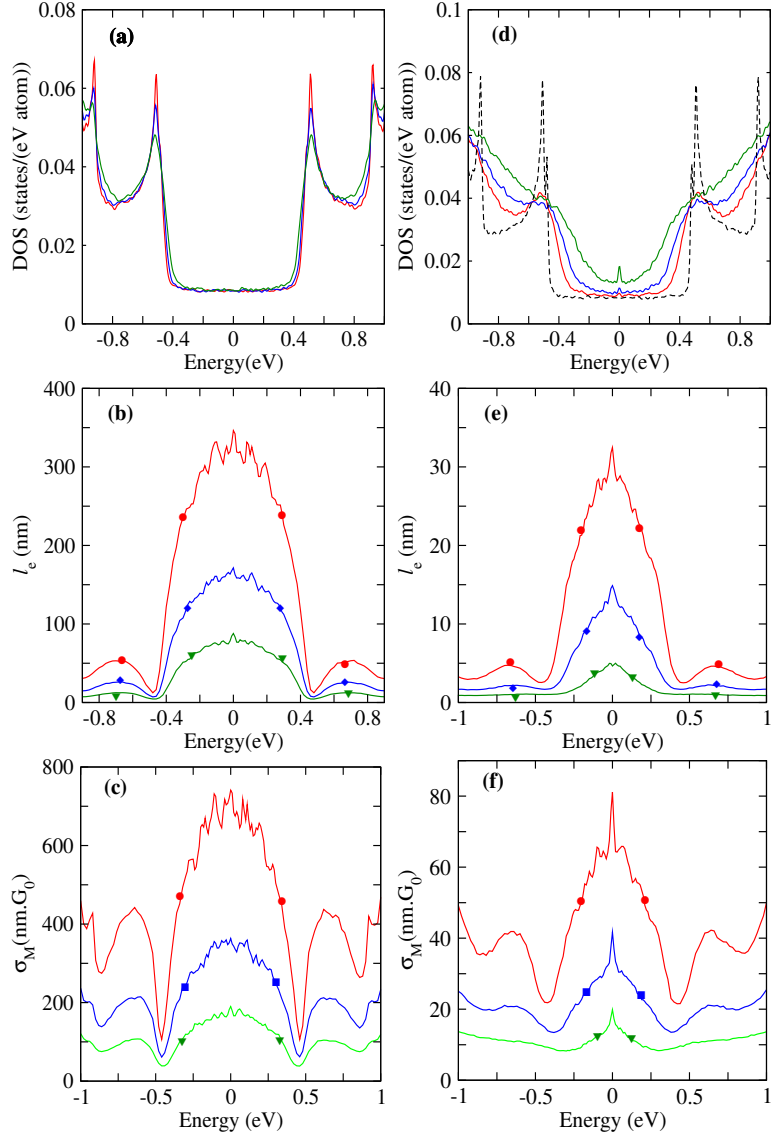

FIG. 3. SWCNT (30.0) with low concentration of non-resonant adsorbates: (a) Density of states DOS(E), (b) Elastic mean-free path $l_e(E)$, and (c) microscopic conductivity $\sigma_M(E)$. SWCNT (30.0) with high concentration of non-resonant adsorbates: (d) DOS(E), (e) $l_e(E)$, and (f) $\sigma_M(E)$.

few of them to those shown in the main text for SWCNT (60,0), in order to discuss the size effect on electronic behavior. Figure 3 shows the Density of states, the elastic mean-free path and the microscopic conductivity for low and high concentration of non-resonant adsorbates. The general behavior is similar to the SWCNT (60,0) already discussed in the main text. In the DOS around the charge neutrality points (CNP) the energy range of the zone where there are only two conductive channels is about 0.8 eV larger than what we observed in SWCNT (60,0) (around 0.4 eV). The values of elastic mean-free path and

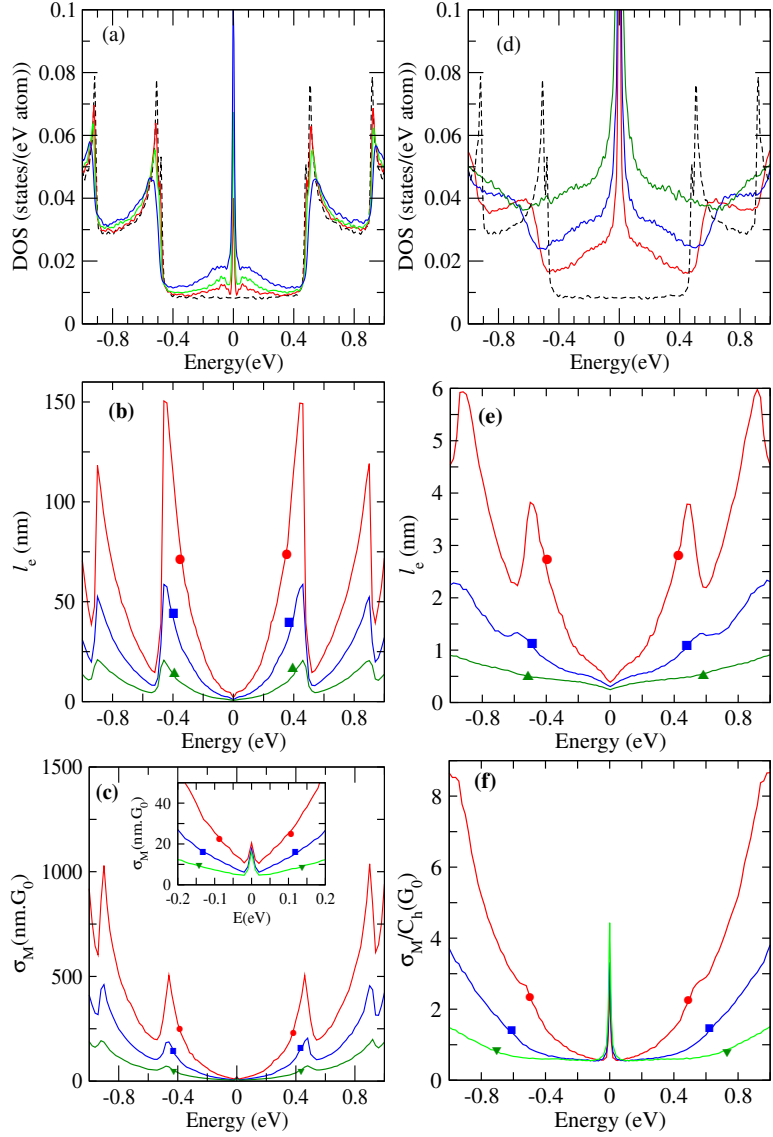

FIG. 4. SWCNT (30.0) with low concentration of resonant adsorbates: (a) Density of states DOS(E), (b) Elastic mean-free path $l_e(E)$, and (c) microscopic conductivity $\sigma_M(E)$. SWCNT (30.0) with high concentration of resonant adsorbates: (d) DOS(E), (e) $l_e(E)$, and (f) $\sigma_M(E)$.

microscopic conductivity are half the values for SWCNT (60,0), for instance, around CNP for 0.1% of non-resonant adsorbates in SWCNT (30,0) the maximum value reached for elastic mean-free path is $l_e = 300$ nm where it is 600 nm for SWCNT (60,0), and $l_e > C_h$ even for very high concentration of non-resonant adsorbates (1%, 2% and 4%). The resonant adsorbates are shown in figure 4, where we can see the pic of resonant states in the DOS around the CNP. By increasing the adsorbates concentration (up to 1%, 2% and 4%) the elastic mean-free path decreases and $l_e < C_h$, so we need a further investigation of the

6

localization length to confirms the possibility of a transition between 1D and 2D electronic behavior. The electronic conductivity versus inelastic mean-free path $\sigma(L_i)$ shown in figure 5-a at energy $E = 0.75$ eV and for low concentration of resonant adsorbates (0.1%, 0.2% and 0.4%) obeys the 1D scale dependent conductivity for large $L_i$. Here, as we did in the main text, we avoid the energies close to the resonant states, we show in figure 5-b, that for an energy $E = 0.3$ eV close to the resonant states the behavior of $\sigma(L_i)$ for large $L_i$ can not be predicted by the 1D scale dependent conductivity because of the strong effects of resonant adsorbates. We show the Thouless relation for the SWCNT (30,0) in figure 5-c and the effective number of conductive channel confirming the 1D behavior in SWCNT (30,0).

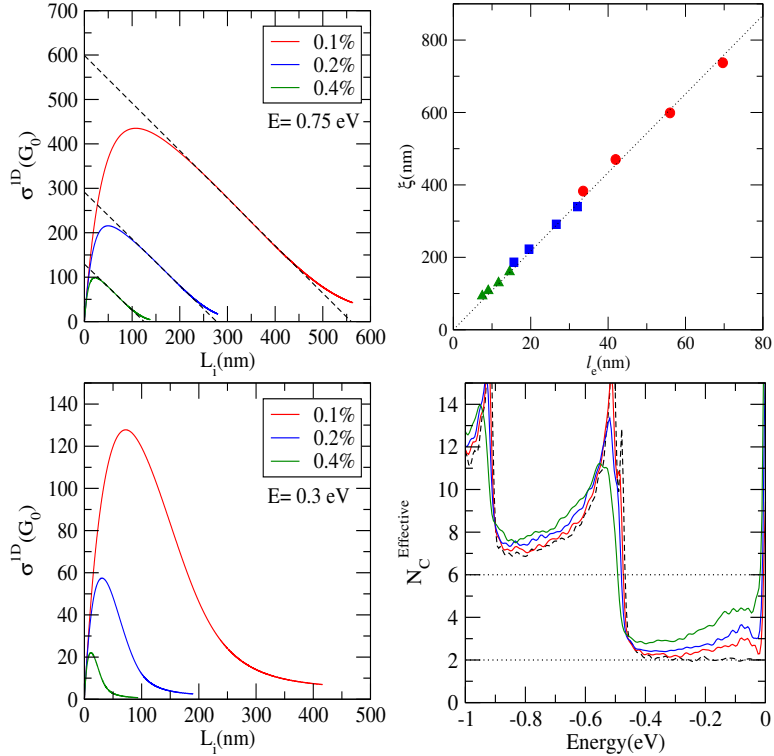

FIG. 5. SWCNT (30.0) with resonant adsorbates: (Left column) Electronic conductivity versus inelastic mean-free path $\sigma(L_i)$ at energy $E = 0.75$ eV, and $\sigma(L_i)$ at $E = 0.3$ eV. (Right column) Thouless relation ($\xi$ versus $l_e$) for the first plateau of energy around $0.75$ eV, and the effective number of conductive channel $N_C^{eff}$ versus energy $N_C^{eff} = n(E)/n_0$, where $n(E)$ is the total density of states and $n_0$ is the density of state for one conductive channel.

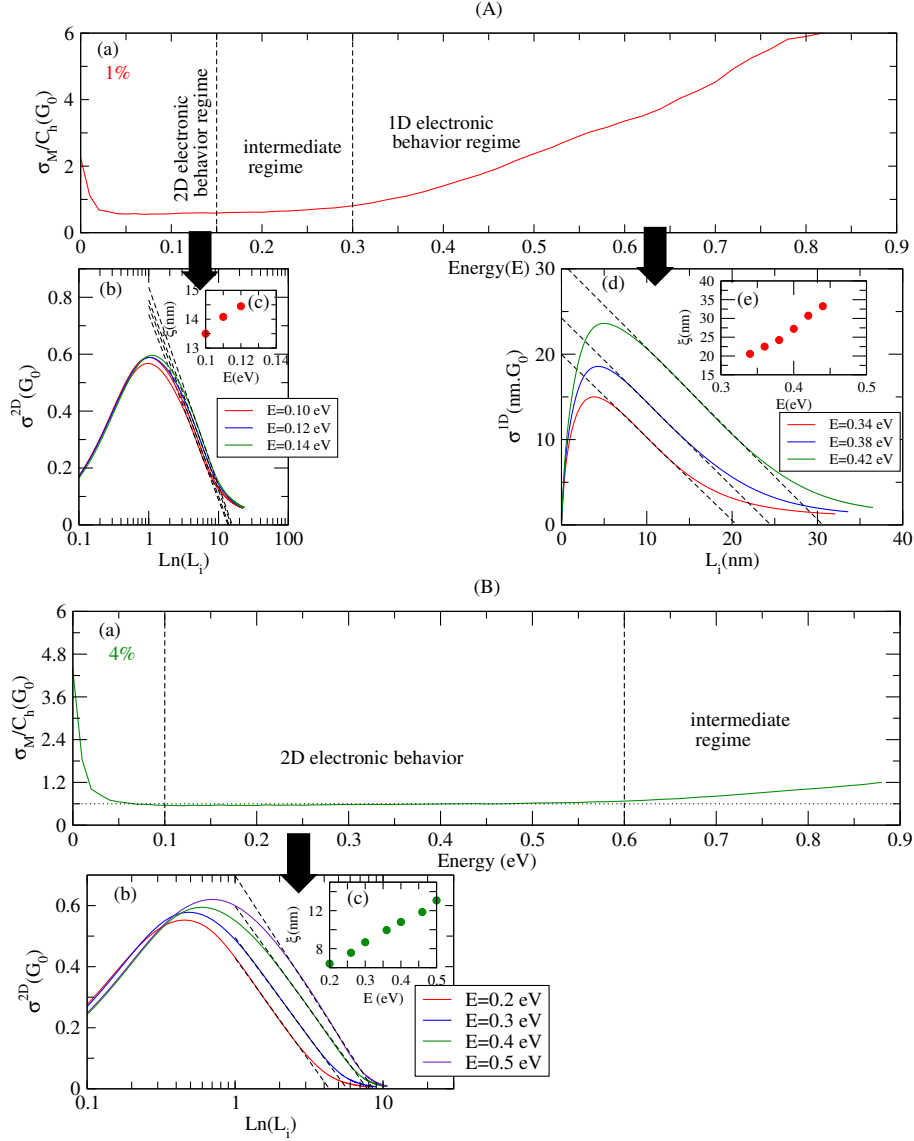

FIG. 6. SWCNT (60,0)(A)1% of resonant adsorbates (B) 4% of resonant adsorbates

## III. THE 1D/2D TRANSITION IN ELECTRONIC BEHABIOR

As discussed in the main text the 2D electronic behavior appears in SWCNT (60,0) at 1% of resonant adsorbates (see figure 6-A ) for a very narrow energy intervall. This energy intervall corresponding to the 2D electronic regime will increases when increasing the resonant adsorbates concentrations until it becomes dominant at 4% of resonant adsorbates as shown in figure 6-B.

However even with 2% of resonant adsorbates in SWCNT (30,0), as shown in figure 7, the 1D regime still dominant and the 2D regime can not be reached, close to resonant adsorbate

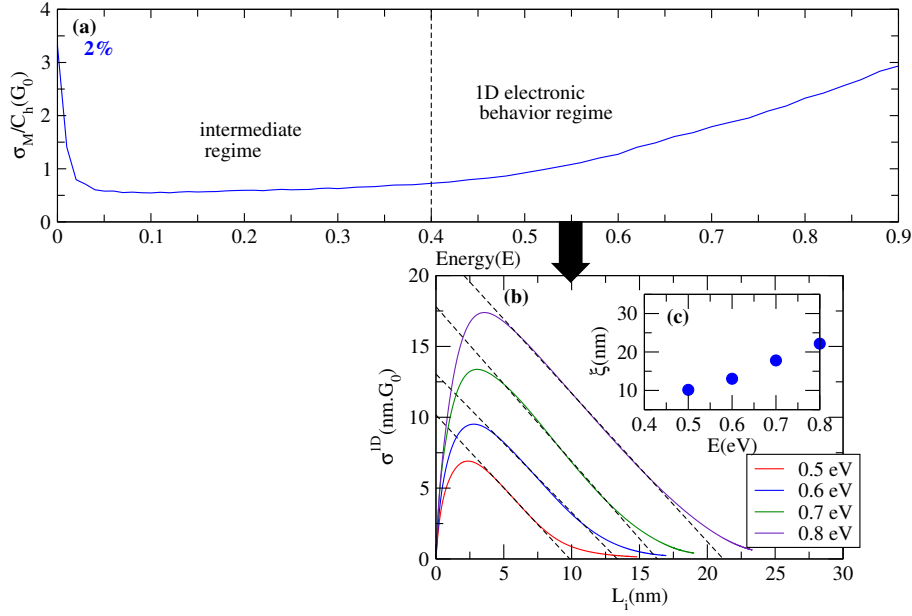

FIG. 7. SWCNT (30.0) 2% of resonant adsorbates

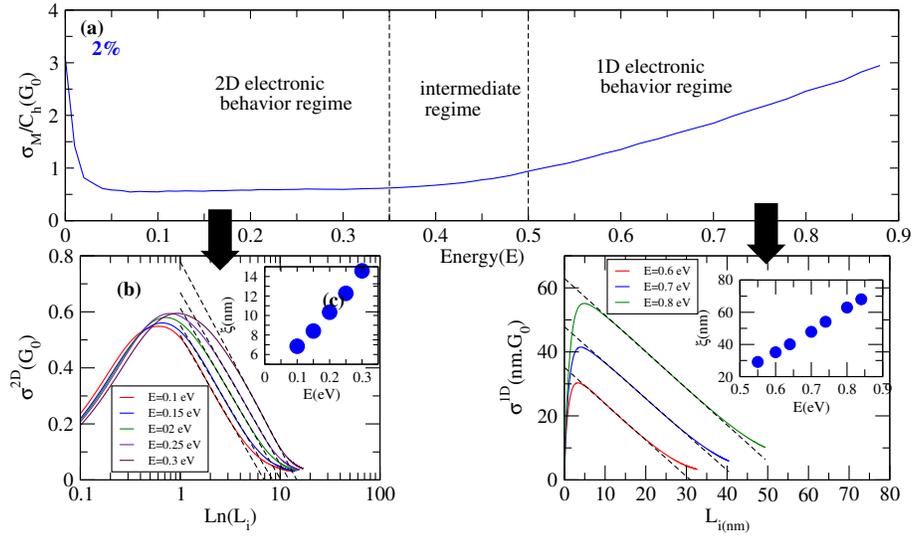

FIG. 8. SWCNT (90.0) 2% of resonant adsorbates

we define an intermediate regime where neither the 1D or 2D scale dependent conductivity can not be applied. In the other hand, reaching the 2D regime is much easier for a larger nanotube such as SWCNT (90,0) (see figure 8). The 2D regime is reached when $\xi < C_h$, and the localization length $\xi$ depends on the size of carbon nanotubes: $\xi$ increases when

the nanotube diameter increases.

---



\end{document}